

\catcode`@=11 

\font\seventeenrm=cmr17
\font\fourteenrm=cmr10 scaled\magstep2
\font\twelverm=cmr12

\font\ninerm=cmr9

\font\sixrm=cmr6

\font\fourteenbf=cmbx10 scaled\magstep2
\font\twelvebf=cmbx10 scaled\magstep1
\font\ninebf=cmbx9      \font\sixbf=cmbx6
\font\seventeeni=cmmi10 scaled\magstep3    \skewchar\seventeeni='177
\font\fourteeni=cmmi10 scaled\magstep2    \skewchar\fourteeni='177
\font\twelvei=cmmi10 scaled\magstep1    \skewchar\twelvei='177
\font\ninei=cmmi9    \skewchar\ninei='177
\font\sixi=cmmi6    \skewchar\sixi='177
\font\seventeensy=cmsy10 scaled\magstep3    \skewchar\seventeensy='60
\font\fourteensy=cmsy10 scaled\magstep2    \skewchar\fourteensy='60
\font\twelvesy=cmsy10 scaled\magstep1    \skewchar\twelvesy='60
\font\ninesy=cmsy9    \skewchar\ninesy='60
\font\sixsy=cmsy6    \skewchar\sixsy='60

\font\fourteenex=cmex10 scaled\magstep2
\font\twelveex=cmex10 scaled\magstep1

\font\fourteensl=cmsl10 scaled\magstep2
\font\twelvesl=cmsl10 scaled\magstep1

\font\ninesl=cmsl9

\font\fourteenit=cmti10 scaled\magstep2
\font\twelveit=cmti10 scaled\magstep1
\font\twelvett=cmtt10 scaled\magstep1

\font\twelvecp=cmcsc10 scaled\magstep1
\font\tencp=cmcsc10
\newfam\cpfam
%
%
\newcount\f@ntkey     \f@ntkey=0
\def\samef@nt{\relax \ifcase\f@ntkey \rm \or\oldstyle \or\or
 \or\it \or\sl \or\bf \or\tt \or\caps \fi }
\def\fourteenpoint{\relax%
    \textfont0=\fourteenrm\scriptfont0=\tenrm%
    \scriptscriptfont0=\sevenrm%
     \def\rm{\fam0\fourteenrm\f@ntkey=0}%
    \textfont1=\fourteeni\scriptfont1=\teni%
    \scriptscriptfont1=\seveni%
     \def\oldstyle{\fam1\fourteeni\f@ntkey=1}%
    \textfont2=\fourteensy\scriptfont2=\tensy%
    \scriptscriptfont2=\sevensy%
    \textfont3=\fourteenex\scriptfont3=\fourteenex%
    \scriptscriptfont3=\fourteenex%
    \def\it{\fam\itfam\fourteenit\f@ntkey=4}\textfont\itfam=\fourteenit%
    \def\sl{\fam\slfam\fourteensl\f@ntkey=5}\textfont\slfam=\fourteensl%
    \scriptfont\slfam=\tensl%
    \def\bf{\fam\bffam\fourteenbf\f@ntkey=6}\textfont\bffam=\fourteenbf%
    \scriptfont\bffam=\tenbf\scriptscriptfont\bffam=\sevenbf%
    \def\tt{\fam\ttfam\twelvett\f@ntkey=7}\textfont\ttfam=\twelvett%
    \h@big=11.9\p@{} \h@Big=16.1\p@{} \h@bigg=20.3\p@{} \h@Bigg=24.5\p@{}%
    \def\caps{\fam\cpfam \twelvecp \f@ntkey=8 }\textfont\cpfam=\twelvecp%
    \setbox\strutbox=\hbox{\vrule height 12pt depth 5pt width\z@}%
    \samef@nt}
\def\twelvepoint{\relax%
    \textfont0=\twelverm\scriptfont0=\ninerm%
    \scriptscriptfont0=\sixrm%
     \def\rm{\fam0\twelverm\f@ntkey=0}%
    \textfont1=\twelvei\scriptfont1=\ninei%
    \scriptscriptfont1=\sixi%
     \def\oldstyle{\fam1\twelvei\f@ntkey=1}%
    \textfont2=\twelvesy\scriptfont2=\ninesy%
    \scriptscriptfont2=\sixsy%
    \textfont3=\twelveex\scriptfont3=\twelveex%
    \scriptscriptfont3=\twelveex%
    \def\it{\fam\itfam\twelveit\f@ntkey=4}\textfont\itfam=\twelveit%
    \def\sl{\fam\slfam\twelvesl\f@ntkey=5}\textfont\slfam=\twelvesl%
    \scriptfont\slfam=\ninesl%
    \def\bf{\fam\bffam\twelvebf\f@ntkey=6}\textfont\bffam=\twelvebf%
    \scriptfont\bffam=\ninebf\scriptscriptfont\bffam=\sixbf%
    \def\tt{\fam\ttfam\twelvett\f@ntkey=7}\textfont\ttfam=\twelvett%
    \h@big=10.2\p@{}%
    \h@Big=13.8\p@{}%
    \h@bigg=17.4\p@{}%
    \h@Bigg=21.0\p@{}%
    \def\caps{\fam\cpfam\twelvecp\f@ntkey=8}\textfont\cpfam=\twelvecp%
    \setbox\strutbox=\hbox{\vrule height 10pt depth 4pt width\z@}%
    \samef@nt}
%
%
\newdimen\h@big  \h@big=8.5\p@
\newdimen\h@Big  \h@Big=11.5\p@
\newdimen\h@bigg  \h@bigg=14.5\p@
\newdimen\h@Bigg  \h@Bigg=17.5\p@
\def\big#1{{\hbox{$\left#1\vbox to\h@big{}\right.\n@space$}}}
\def\Big#1{{\hbox{$\left#1\vbox to\h@Big{}\right.\n@space$}}}
\def\bigg#1{{\hbox{$\left#1\vbox to\h@bigg{}\right.\n@space$}}}
\def\Bigg#1{{\hbox{$\left#1\vbox to\h@Bigg{}\right.\n@space$}}}
%
%
\normalbaselineskip = 20pt plus 0.2pt minus 0.1pt
\normallineskip = 1.5pt plus 0.1pt minus 0.1pt
\normallineskiplimit = 1.5pt
\newskip\normaldisplayskip
\normaldisplayskip = 20pt plus 5pt minus 10pt
\newskip\normaldispshortskip
\normaldispshortskip = 6pt plus 5pt
\newskip\normalparskip
\normalparskip = 6pt plus 2pt minus 1pt
\newskip\skipregister
\skipregister = 5pt plus 2pt minus 1.5pt
\newif\ifsingl@   \newif\ifdoubl@
\newif\iftwelv@   \twelv@true
\def\singlespace{\singl@true\doubl@false\spaces@t}
\def\doublespace{\singl@false\doubl@true\spaces@t}
\def\normalspace{\singl@false\doubl@false\spaces@t}
\def\Tenpoint{\tenpoint\twelv@false\spaces@t}
\def\Twelvepoint{\twelvepoint\twelv@true\spaces@t}
\def\spaces@t{\relax%
 \iftwelv@ \ifsingl@\subspaces@t3:4;\else\subspaces@t1:1;\fi%
 \else \ifsingl@\subspaces@t3:5;\else\subspaces@t4:5;\fi \fi%
 \ifdoubl@ \multiply\baselineskip by 5%
 \divide\baselineskip by 4 \fi \unskip}
\def\subspaces@t#1:#2;{%
      \baselineskip = \normalbaselineskip%
      \multiply\baselineskip by #1 \divide\baselineskip by #2%
      \lineskip = \normallineskip%
      \multiply\lineskip by #1 \divide\lineskip by #2%
      \lineskiplimit = \normallineskiplimit%
      \multiply\lineskiplimit by #1 \divide\lineskiplimit by #2%
      \parskip = \normalparskip%
      \multiply\parskip by #1 \divide\parskip by #2%
      \abovedisplayskip = \normaldisplayskip%
      \multiply\abovedisplayskip by #1 \divide\abovedisplayskip by #2%
      \belowdisplayskip = \abovedisplayskip%
      \abovedisplayshortskip = \normaldispshortskip%
      \multiply\abovedisplayshortskip by #1%
 \divide\abovedisplayshortskip by #2%
      \belowdisplayshortskip = \abovedisplayshortskip%
      \advance\belowdisplayshortskip by \belowdisplayskip%
      \divide\belowdisplayshortskip by 2%
      \smallskipamount = \skipregister%
      \multiply\smallskipamount by #1 \divide\smallskipamount by #2%
      \medskipamount = \smallskipamount \multiply\medskipamount by 2%
      \bigskipamount = \smallskipamount \multiply\bigskipamount by 4 }
\def\normalbaselines{ \baselineskip=\normalbaselineskip%
   \lineskip=\normallineskip \lineskiplimit=\normallineskip%
   \iftwelv@\else \multiply\baselineskip by 4 \divide\baselineskip by 5%
     \multiply\lineskiplimit by 4 \divide\lineskiplimit by 5%
     \multiply\lineskip by 4 \divide\lineskip by 5 \fi }
\Twelvepoint  
\interlinepenalty=50
\interfootnotelinepenalty=5000
\predisplaypenalty=9000
\postdisplaypenalty=500
\hfuzz=1pt
\vfuzz=0.2pt
%
%
\def\pagecontents{%
   \ifvoid\topins\else\unvbox\topins\vskip\skip\topins\fi
   \dimen@ = \dp255 \unvbox255
   \ifvoid\footins\else\vskip\skip\footins\footrule\unvbox\footins\fi
   \ifr@ggedbottom \kern-\dimen@ \vfil \fi }
\def\makeheadline{\vbox to 0pt{ \skip@=\topskip
      \advance\skip@ by -12pt \advance\skip@ by -2\normalbaselineskip
      \vskip\skip@ \line{\vbox to 12pt{}\the\headline} \vss
      }\nointerlineskip}
\def\makefootline{\baselineskip = 1.5\normalbaselineskip
 \line{\the\footline}}
\newif\iffrontpage
\newif\ifletterstyle
\newif\ifp@genum
\def\nopagenumbers{\p@genumfalse}
\def\pagenumbers{\p@genumtrue}
\pagenumbers
\newtoks\paperheadline
\newtoks\letterheadline
\newtoks\letterfrontheadline
\newtoks\lettermainheadline
\newtoks\paperfootline
\newtoks\letterfootline
\newtoks\date
\footline={\ifletterstyle\the\letterfootline\else\the\paperfootline\fi}
\paperfootline={\hss\iffrontpage\else\ifp@genum\tenrm
    -- \folio\ --\hss\fi\fi}
\letterfootline={\hfil}
\headline={\ifletterstyle\the\letterheadline\else\the\paperheadline\fi}
\paperheadline={\hfil}
\def\monthname{\relax\ifcase\month 0/\or January\or February\or
   March\or April\or May\or June\or July\or August\or September\or
   October\or November\or December\else\number\month/\fi}
\date={\monthname\ \number\day, \number\year}
\countdef\pagenumber=1  \pagenumber=1
\def\advancepageno{\global\advance\pageno by 1
   \ifnum\pagenumber<0 \global\advance\pagenumber by -1
    \else\global\advance\pagenumber by 1 \fi \global\frontpagefalse }
\def\folio{\ifnum\pagenumber<0 \romannumeral-\pagenumber
   \else \number\pagenumber \fi }
\def\footrule{\dimen@=\prevdepth\nointerlineskip
   \vbox to 0pt{\vskip -0.25\baselineskip \hrule width 0.35\hsize \vss}
   \prevdepth=\dimen@ }
\newtoks\foottokens
\foottokens={\Tenpoint\singlespace}
\newdimen\footindent
\footindent=24pt
\def\vfootnote#1{\insert\footins\bgroup  \the\foottokens
   \interlinepenalty=\interfootnotelinepenalty \floatingpenalty=20000
   \splittopskip=\ht\strutbox \boxmaxdepth=\dp\strutbox
   \leftskip=\footindent \rightskip=\z@skip
   \parindent=0.5\footindent \parfillskip=0pt plus 1fil
   \spaceskip=\z@skip \xspaceskip=\z@skip
   \Textindent{$ #1 $}\footstrut\futurelet\next\fo@t}
\def\Textindent#1{\noindent\llap{#1\enspace}\ignorespaces}
\def\footnote#1{\attach{#1}\vfootnote{#1}}

\newcount\lastf@@t     \lastf@@t=-1
\newcount\footsymbolcount    \footsymbolcount=0
\newif\ifPhysRev
\def\space@ver#1{\let\@sf=\empty \ifmmode #1\else \ifhmode
   \edef\@sf{\spacefactor=\the\spacefactor}\unskip${}#1$\relax\fi\fi}
\def\attach#1{\space@ver{\strut^{\mkern 2mu #1} }\@sf\ }
%
%
%
\newcount\chapternumber     \chapternumber=0
\newcount\sectionnumber     \sectionnumber=0
\newcount\equanumber     \equanumber=0
\let\chapterlabel=0
\newtoks\chapterstyle     \chapterstyle={\Number}
\newskip\chapterskip     \chapterskip=\bigskipamount
\newskip\sectionskip     \sectionskip=\medskipamount
\newskip\headskip     \headskip=8pt plus 3pt minus 3pt
\newdimen\chapterminspace    \chapterminspace=15pc
\newdimen\sectionminspace    \sectionminspace=10pc
\newdimen\referenceminspace  \referenceminspace=25pc
\def\chapterreset{\global\advance\chapternumber by 1
   \ifnum\equanumber<0 \else\global\equanumber=0\fi
   \sectionnumber=0 \makel@bel}
\def\makel@bel{\xdef\chapterlabel{%
\the\chapterstyle{\the\chapternumber}.}}
\def\sectionlabel{\number\sectionnumber \quad }
\def\alphabetic#1{\count255='140 \advance\count255 by #1\char\count255}
\def\Alphabetic#1{\count255='100 \advance\count255 by #1\char\count255}
\def\Roman#1{\uppercase\expandafter{\romannumeral #1}}
\def\roman#1{\romannumeral #1}
\def\Number#1{\number #1}
\def\unnumberedchapters{\let\makel@bel=\relax \let\chapterlabel=\relax
\let\sectionlabel=\relax \equanumber=-1 }
%
\def\titlestyle#1{\par\begingroup \interlinepenalty=9999
     \leftskip=0.03\hsize plus 0.20\hsize minus 0.03\hsize
     \rightskip=\leftskip \parfillskip=0pt
     \hyphenpenalty=9000 \exhyphenpenalty=9000
     \tolerance=9999 \pretolerance=9000
     \spaceskip=0.333em \xspaceskip=0.5em
     \iftwelv@\fourteenpoint\fourteenbf\else\twelvepoint\twelvebf\fi
     \noindent  #1\par\endgroup }

\def\spacecheck#1{\dimen@=\pagegoal\advance\dimen@ by -\pagetotal
   \ifdim\dimen@<#1 \ifdim\dimen@>0pt \vfil\break \fi\fi}
\def\chapter#1{\par \penalty-300 \vskip\chapterskip
   \spacecheck\chapterminspace
   \chapterreset \titlestyle{\chapterlabel \ #1}
   \nobreak\vskip\headskip \penalty 30000
   \wlog{\string\chapter\ \chapterlabel} }
%
\def\section#1{\par \ifnum\the\lastpenalty=30000\else
   \penalty-200\vskip\sectionskip \spacecheck\sectionminspace\fi
   \wlog{\string\section\ \chapterlabel \the\sectionnumber}
   \global\advance\sectionnumber by 1  \noindent
   {\caps\enspace\chapterlabel \sectionlabel #1}\par
   \nobreak\vskip\headskip \penalty 30000 }
\def\subsection#1{\par
   \ifnum\the\lastpenalty=30000\else \penalty-100\smallskip \fi
   \noindent\undertext{#1}\enspace \vadjust{\penalty5000}}

\def\undertext#1{\vtop{\hbox{#1}\kern 1pt \hrule}}
\def\ack{\par\penalty-100\medskip \spacecheck\sectionminspace
   \line{\fourteenrm\hfil ACKNOWLEDGEMENTS\hfil}\nobreak\vskip\headskip }
\def\APPENDIX#1#2{\par\penalty-300\vskip\chapterskip
   \spacecheck\chapterminspace \chapterreset \xdef\chapterlabel{#1}
   \titlestyle{APPENDIX #2} \nobreak\vskip\headskip \penalty 30000
   \wlog{\string\Appendix\ \chapterlabel} }
\def\Appendix#1{\APPENDIX{#1}{#1}}
\def\appendix{\APPENDIX{A}{}}
%
%
%
\def\eqname#1{\relax \ifnum\equanumber<0
     \xdef#1{{\rm(\number-\equanumber)}}\global\advance\equanumber by -1
    \else \global\advance\equanumber by 1
      \xdef#1{{\rm(\chapterlabel \number\equanumber)}} \fi}
\def\eq{\eqname\?\?}
\def\eqn#1{\eqno\eqname{#1}#1}

\def\eqinsert#1{\noalign{\dimen@=\prevdepth \nointerlineskip
   \setbox0=\hbox to\displaywidth{\hfil #1}
   \vbox to 0pt{\vss\hbox{$\!\box0\!$}\kern-0.5\baselineskip}
   \prevdepth=\dimen@}}
%

%

%

%
%
\def\GENITEM#1;#2{\par \hangafter=0 \hangindent=#1
    \Textindent{#2}\ignorespaces}
\outer\def\newitem#1=#2;{\gdef#1{\GENITEM #2;}}
\newdimen\itemsize  \itemsize=30pt
\newitem\item=1\itemsize;
\newitem\sitem=1.75\itemsize;  
\newitem\ssitem=2.5\itemsize;  
\outer\def\newlist#1=#2&#3&#4;{\toks0={#2}\toks1={#3}%
   \count255=\escapechar \escapechar=-1
   \alloc@0\list\countdef\insc@unt\listcount \listcount=0
   \edef#1{\par
      \countdef\listcount=\the\allocationnumber
      \advance\listcount by 1
      \hangafter=0 \hangindent=#4
      \Textindent{\the\toks0{\listcount}\the\toks1}}
   \expandafter\expandafter\expandafter
    \edef\c@t#1{begin}{\par
      \countdef\listcount=\the\allocationnumber \listcount=1
      \hangafter=0 \hangindent=#4
      \Textindent{\the\toks0{\listcount}\the\toks1}}
   \expandafter\expandafter\expandafter
    \edef\c@t#1{con}{\par \hangafter=0 \hangindent=#4 \noindent}
   \escapechar=\count255}
\def\c@t#1#2{\csname\string#1#2\endcsname}
\newlist\point=\Number&.&1.0\itemsize;
\newlist\subpoint=(\alphabetic&)&1.75\itemsize;
\newlist\subsubpoint=(\roman&)&2.5\itemsize;
%

%
%
%
\newcount\referencecount     \referencecount=0
\newif\ifreferenceopen     \newwrite\referencewrite
\newtoks\rw@toks
\def\NPrefmark#1{\attach{\scriptscriptstyle[#1]}}
\let\PRrefmark=\attach
\def\refmark#1{\relax\ifPhysRev\PRrefmark{#1}\else\NPrefmark{#1}\fi}
\def\refend{\refmark{\number\referencecount}}
\newcount\lastrefsbegincount \lastrefsbegincount=0
\def\refsend{\refmark{\count255=\referencecount
   \advance\count255 by-\lastrefsbegincount
   \ifcase\count255 \number\referencecount
   \or \number\lastrefsbegincount,\number\referencecount
   \else \number\lastrefsbegincount-\number\referencecount \fi}}
\def\refch@ck{\chardef\rw@write=\referencewrite
   \ifreferenceopen \else \referenceopentrue
   \immediate\openout\referencewrite=reference.aux \fi}
%
{\catcode`\^^M=\active 
  \gdef\obeyendofline{\catcode`\^^M\active \let^^M\ }}%
%
{\catcode`\^^M=\active 
  \gdef\ignoreendofline{\catcode`\^^M=5}}
{\obeyendofline\gdef\rw@start#1{\def\t@st{#1} \ifx\t@st\blankend%
\endgroup \@sf \relax \else \ifx\t@st\bl@nkend \endgroup \@sf \relax%
\else \rw@begin#1
\backtotext
\fi \fi } }
{\obeyendofline\gdef\rw@begin#1
{\def\n@xt{#1}\rw@toks={#1}\relax%
\rw@next}}
\def\blankend{}
{\obeylines\gdef\bl@nkend{
}}
\newif\iffirstrefline  \firstreflinetrue
\def\rwr@teswitch{\ifx\n@xt\blankend \let\n@xt=\rw@begin %
 \else\iffirstrefline \global\firstreflinefalse%
\immediate\write\rw@write{\noexpand\obeyendofline \the\rw@toks}%
\let\n@xt=\rw@begin%
      \else\ifx\n@xt\rw@@d \def\n@xt{\immediate\write\rw@write{%
\noexpand\ignoreendofline}\endgroup \@sf}%
     \else \immediate\write\rw@write{\the\rw@toks}%
     \let\n@xt=\rw@begin\fi\fi \fi}
\def\rw@next{\rwr@teswitch\n@xt}
\def\rw@@d{\backtotext} \let\rw@end=\relax
\let\backtotext=\relax

\newdimen\refindent\refindent=30pt
\def\refitem#1{\par \hangafter=0 \hangindent=\refindent \Textindent{#1}}
\def\REFNUM#1{\space@ver{}\refch@ck \firstreflinetrue%
 \global\advance\referencecount by 1 \xdef#1{\the\referencecount}}
\def\refnum#1{\space@ver{}\refch@ck \firstreflinetrue%
 \global\advance\referencecount by 1 \xdef#1{\the\referencecount}\refend}

\def\REF#1{\REFNUM#1%
 \immediate\write\referencewrite{%
 \noexpand\refitem{#1.}}%
\begingroup\obeyendofline\rw@start}
\def\ref{\refnum\?%
 \immediate\write\referencewrite{\noexpand\refitem{\?.}}%
\begingroup\obeyendofline\rw@start}
\def\Ref#1{\refnum#1%
 \immediate\write\referencewrite{\noexpand\refitem{#1.}}%
\begingroup\obeyendofline\rw@start}
\newcount\figurecount  \figurecount=0
\newif\iffigureopen  \newwrite\figurewrite
\def\figch@ck{\chardef\rw@write=\figurewrite \iffigureopen\else
   \immediate\openout\figurewrite=figures.aux
   \figureopentrue\fi}
\def\FIGNUM#1{\space@ver{}\figch@ck \firstreflinetrue%
 \global\advance\figurecount by 1 \xdef#1{\the\figurecount}}
\def\FIG#1{\FIGNUM#1
   \immediate\write\figurewrite{\noexpand\refitem{#1.}}%
   \begingroup\obeyendofline\rw@start}
\def\fig{\FIGNUM\? fig.~\?%
\immediate\write\figurewrite{\noexpand\refitem{\?.}}%
\begingroup\obeyendofline\rw@start}
\def\figure{\FIGNUM\? figure~\?
   \immediate\write\figurewrite{\noexpand\refitem{\?.}}%
   \begingroup\obeyendofline\rw@start}
\def\Fig{\FIGNUM\? Fig.~\?%
\immediate\write\figurewrite{\noexpand\refitem{\?.}}%
\begingroup\obeyendofline\rw@start}
\def\Figure{\FIGNUM\? Figure~\?%
\immediate\write\figurewrite{\noexpand\refitem{\?.}}%
\begingroup\obeyendofline\rw@start}
\newcount\tablecount \tablecount=0
\newif\iftableopen \newwrite\tablewrite
\def\tabch@ck{\chardef\rw@write=\tablewrite \iftableopen\else
   \immediate\openout\tablewrite=tables.aux
   \tableopentrue\fi}
\def\TABNUM#1{\space@ver{}\tabch@ck \firstreflinetrue%
 \global\advance\tablecount by 1 \xdef#1{\the\tablecount}}
\def\TABLE#1{\TABNUM#1
   \immediate\write\tablewrite{\noexpand\refitem{#1.}}%
   \begingroup\obeyendofline\rw@start}
\def\Table{\TABNUM\? Table~\?%
\immediate\write\tablewrite{\noexpand\refitem{\?.}}%
\begingroup\obeyendofline\rw@start}
%

%
%
%
\def\masterreset{\global\pagenumber=1 \global\chapternumber=0
   \global\equanumber=0 \global\sectionnumber=0
   \global\referencecount=0 \global\figurecount=0 \global\tablecount=0 }
\def\FRONTPAGE{\ifvoid255\else\vfill\penalty-2000\fi
      \masterreset\global\frontpagetrue
      \global\lastf@@t=0 \global\footsymbolcount=0}

\def\paperstyle{\letterstylefalse\normalspace\papersize}
\def\papersize{\hsize=35.2pc\vsize=52.7pc\hoffset=0.5pc\voffset=0.8pc
       \skip\footins=\bigskipamount}
\paperstyle   
\newskip\spskip \setbox0\hbox{\ } \spskip=-1\wd0
\def\addressee#1{\medskip\rightline{\the\date\hskip 30pt} \bigskip
   \vskip\lettertopfil
   \ialign to\hsize{\strut ##\hfil\tabskip 0pt plus \hsize \cr #1\crcr}
   \medskip\noindent\hskip\spskip}
\newskip\signatureskip     \signatureskip=40pt
\def\signed#1{\par \penalty 9000 \bigskip \dt@pfalse
  \everycr={\noalign{\ifdt@p\vskip\signatureskip\global\dt@pfalse\fi}}
  \setbox0=\vbox{\singlespace \halign{\tabskip 0pt \strut ##\hfil\cr
   \noalign{\global\dt@ptrue}#1\crcr}}
  \line{\hskip 0.5\hsize minus 0.5\hsize \box0\hfil} \medskip }

\def\endletter{\ifnum\pagenumber=1 \vskip\letterbottomfil\supereject
\else \vfil\supereject \fi}
\newbox\letterb@x
\def\lettertext{\par\unvcopy\letterb@x\par}
\def\multiletter{\setbox\letterb@x=\vbox\bgroup
      \everypar{\vrule height 1\baselineskip depth 0pt width 0pt }
      \singlespace \topskip=\baselineskip }
\def\letterend{\par\egroup}
%
%
%
\newskip\frontpageskip
\newtoks\pubtype
\newtoks\Pubnum
\newtoks\pubnum
\newif\ifp@bblock \p@bblocktrue
\def\nopubblock{\p@bblockfalse}
\def\endpage{\vfil\break}
\frontpageskip=1\medskipamount plus .5fil
\pubtype={ }
\newtoks\publevel
\publevel={Report}   
\Pubnum={\the\pubnum}
\def\p@bblock{\begingroup \tabskip=\hsize minus \hsize
   \baselineskip=1.5\ht\strutbox \topspace-2\baselineskip
   \halign to\hsize{\strut ##\hfil\tabskip=0pt\crcr
   \the\Pubnum\cr \the\date\cr }\endgroup}
\def\title#1{\vskip\frontpageskip\vfill
   {\fourteenbf\titlestyle{#1}}\vskip\headskip\vfill }
\def\author#1{\vskip\frontpageskip\titlestyle{\twelvecp #1}\nobreak}

\def\address#1{\par\kern 5pt \titlestyle{\twelvepoint\sl #1}}
\def\abstract#1{\vfill\vskip\frontpageskip\centerline{\fourteenrm ABSTRACT}
                \vskip\headskip#1\endpage}
%

%
\def\ie{\hbox{\it i.e.}}

\def\\{\relax\ifmmode\backslash\else$\backslash$\fi}
\def\globaleqnumbers{\relax\if\equanumber<0\else\global\equanumber=-1\fi}
\def\nextline{\unskip\nobreak\hskip\parfillskip\break}
\def\subpar{\vadjust{\allowbreak\vskip\parskip}\nextline}
\def\journal#1&#2(#3){\unskip, \sl #1~\bf #2 \rm (19#3) }
\def\cropen#1{\crcr\noalign{\vskip #1}}
\def\crr{\cropen{10pt}}
\def\topspace{\hrule height 0pt depth 0pt \vskip}

\def\VEV#1{\left\langle #1\right\rangle}
\def\Tr{\mathop{\rm Tr}\nolimits}
\let\int=\intop
\def\prop{\mathrel{{\mathchoice{\pr@p\scriptstyle}{\pr@p\scriptstyle}{
\pr@p\scriptscriptstyle}{\pr@p\scriptscriptstyle} }}}
\def\pr@p#1{\setbox0=\hbox{$\cal #1 \char'103$}
   \hbox{$\cal #1 \char'117$\kern-.4\wd0\box0}}
\def\lsim{\mathrel{\mathpalette\@versim<}}
\def\gsim{\mathrel{\mathpalette\@versim>}}
\def\@versim#1#2{\lower0.2ex\vbox{\baselineskip\z@skip\lineskip\z@skip
  \lineskiplimit\z@\ialign{$\m@th#1\hfil##\hfil$\crcr#2\crcr\sim\crcr}}}
%
%
%
\let\sec@nt=\sec
\def\sec{\relax\ifmmode\let\n@xt=\sec@nt\else\let\n@xt\section\fi\n@xt}
\def\obsolete#1{\message{Macro \string #1 is obsolete.}}
\def\firstsec#1{\obsolete\firstsec \section{#1}}
\def\firstsubsec#1{\obsolete\firstsubsec \subsection{#1}}
\def\thispage#1{\obsolete\thispage \global\pagenumber=#1\frontpagefalse}
\def\thischapter#1{\obsolete\thischapter \global\chapternumber=#1}
\def\nextequation#1{\obsolete\nextequation \global\equanumber=#1
   \ifnum\the\equanumber>0 \global\advance\equanumber by 1 \fi}
\def\BOXITEM{\afterassigment\B@XITEM\setbox0=}
\def\B@XITEM{\par\hangindent\wd0 \noindent\box0 }
%

%
\catcode`@=12 
\message{phyzzx.tex by V.K.}
\relax

%
\catcode`@=11
%
%
\font\fourteenmib=cmmib10 scaled\magstep2    \skewchar\fourteenmib='177
\font\twelvemib=cmmib10 scaled\magstep1    \skewchar\twelvemib='177
\font\elevenmib=cmmib10 scaled\magstephalf   \skewchar\elevenmib='177
\font\tenmib=cmmib10    \skewchar\tenmib='177
\font\fourteenbsy=cmbsy10 scaled\magstep2     \skewchar\fourteenbsy='60
\font\twelvebsy=cmbsy10 scaled\magstep1      \skewchar\twelvebsy='60
\font\elevenbsy=cmbsy10 scaled\magstephalf    \skewchar\elevenbsy='60
\font\tenbsy=cmbsy10      \skewchar\tenbsy='60
\newfam\mibfam
\def\samef@nt{\relax \ifcase\f@ntkey \rm \or\oldstyle \or\or
 \or\it \or\sl \or\bf \or\tt \or\caps \or\mib \fi }
\def\fourteenpoint{\relax
    \textfont0=\fourteenrm    \scriptfont0=\tenrm
    \scriptscriptfont0=\sevenrm
     \def\rm{\fam0 \fourteenrm \f@ntkey=0 }\relax
    \textfont1=\fourteeni    \scriptfont1=\teni
    \scriptscriptfont1=\seveni
     \def\oldstyle{\fam1 \fourteeni\f@ntkey=1 }\relax
    \textfont2=\fourteensy    \scriptfont2=\tensy
    \scriptscriptfont2=\sevensy
    \textfont3=\fourteenex     \scriptfont3=\fourteenex
    \scriptscriptfont3=\fourteenex
    \def\it{\fam\itfam \fourteenit\f@ntkey=4 }\textfont\itfam=\fourteenit
    \def\sl{\fam\slfam \fourteensl\f@ntkey=5 }\textfont\slfam=\fourteensl
    \scriptfont\slfam=\tensl
    \def\bf{\fam\bffam \fourteenbf\f@ntkey=6 }\textfont\bffam=\fourteenbf
    \scriptfont\bffam=\tenbf \scriptscriptfont\bffam=\sevenbf
    \def\tt{\fam\ttfam \twelvett \f@ntkey=7 }\textfont\ttfam=\twelvett
    \h@big=11.9\p@{} \h@Big=16.1\p@{} \h@bigg=20.3\p@{} \h@Bigg=24.5\p@{}
    \def\caps{\fam\cpfam \twelvecp \f@ntkey=8 }\textfont\cpfam=\twelvecp
    \setbox\strutbox=\hbox{\vrule height 12pt depth 5pt width\z@}
    \def\mib{\fam\mibfam \fourteenmib \f@ntkey=9 }
    \textfont\mibfam=\fourteenmib      \scriptfont\mibfam=\tenmib
    \scriptscriptfont\mibfam=\tenmib
    \samef@nt}
\def\twelvepoint{\relax
    \textfont0=\twelverm  \scriptfont0=\ninerm
    \scriptscriptfont0=\sixrm
     \def\rm{\fam0 \twelverm \f@ntkey=0 }\relax
    \textfont1=\twelvei  \scriptfont1=\ninei
    \scriptscriptfont1=\sixi
     \def\oldstyle{\fam1 \twelvei\f@ntkey=1 }\relax
    \textfont2=\twelvesy  \scriptfont2=\ninesy
    \scriptscriptfont2=\sixsy
    \textfont3=\twelveex  \scriptfont3=\twelveex
    \scriptscriptfont3=\twelveex
    \def\it{\fam\itfam \twelveit \f@ntkey=4 }\textfont\itfam=\twelveit
    \def\sl{\fam\slfam \twelvesl \f@ntkey=5 }\textfont\slfam=\twelvesl
    \scriptfont\slfam=\ninesl
    \def\bf{\fam\bffam \twelvebf \f@ntkey=6 }\textfont\bffam=\twelvebf
    \scriptfont\bffam=\ninebf  \scriptscriptfont\bffam=\sixbf
    \def\tt{\fam\ttfam \twelvett \f@ntkey=7 }\textfont\ttfam=\twelvett
    \h@big=10.2\p@{}
    \h@Big=13.8\p@{}
    \h@bigg=17.4\p@{}
    \h@Bigg=21.0\p@{}
    \def\caps{\fam\cpfam \twelvecp \f@ntkey=8 }\textfont\cpfam=\twelvecp
    \setbox\strutbox=\hbox{\vrule height 10pt depth 4pt width\z@}
    \def\mib{\fam\mibfam \twelvemib \f@ntkey=9 }
    \textfont\mibfam=\twelvemib    \scriptfont\mibfam=\tenmib
    \scriptscriptfont\mibfam=\tenmib
    \samef@nt}
\def\tenpoint{\relax
    \textfont0=\tenrm       \scriptfont0=\sevenrm
    \scriptscriptfont0=\fiverm
    \def\rm{\fam0 \tenrm \f@ntkey=0 }\relax
    \textfont1=\teni       \scriptfont1=\seveni
    \scriptscriptfont1=\fivei
    \def\oldstyle{\fam1 \teni \f@ntkey=1 }\relax
    \textfont2=\tensy       \scriptfont2=\sevensy
    \scriptscriptfont2=\fivesy
    \textfont3=\tenex       \scriptfont3=\tenex
    \scriptscriptfont3=\tenex
    \def\it{\fam\itfam \tenit \f@ntkey=4 }\textfont\itfam=\tenit
    \def\sl{\fam\slfam \tensl \f@ntkey=5 }\textfont\slfam=\tensl
    \def\bf{\fam\bffam \tenbf \f@ntkey=6 }\textfont\bffam=\tenbf
    \scriptfont\bffam=\sevenbf   \scriptscriptfont\bffam=\fivebf
    \def\tt{\fam\ttfam \tentt \f@ntkey=7 }\textfont\ttfam=\tentt
    \def\caps{\fam\cpfam \tencp \f@ntkey=8 }\textfont\cpfam=\tencp
    \setbox\strutbox=\hbox{\vrule height 8.5pt depth 3.5pt width\z@}
    \def\mib{\fam\mibfam \tenmib \f@ntkey=9 }
    \textfont\mibfam=\tenmib   \scriptfont\mibfam=\tenmib
    \scriptscriptfont\mibfam=\tenmib
    \samef@nt}
%
%
%
%
\Twelvepoint 
\catcode`@=12
\def\unlock{\catcode`@=11} 
\def\lock{\catcode`@=12} 
\unlock
\fontdimen5\textfont2=1.2pt
 \def\ee{\eqno\eq }

\def\KUNSmark{\vtop{\hbox{\elevenmib Department\hskip1mm of\hskip1mm
             Physics}\hbox{\elevenmib Kyoto\hskip1mm University}}}
\newif\ifKUNS \KUNStrue
\def\titlepage{\FRONTPAGE\paperstyle\ifPhysRev\PH@SR@V\fi
    \ifKUNS\null\vskip-17mm\KUNSmark\vskip0mm\fi
    \ifp@bblock\p@bblock\fi}
\Pubnum={\ifdraft\undertext{\strut$\mib draft$}\cr\fi
         KUNS~\the\pubnum}
\newcount\YEAR
\YEAR=\number\year
\global\advance\YEAR by-1900
\pubnum={\number\YEAR-XX}
\newif\ifdraft
\lock
\message{ modified by K-I. A.}

\def\Gab{G_{\alpha\beta}}
\def\Gabu{G^{\alpha\beta}}
\def\nba{\nabla_\alpha}
\def\nbb{\nabla_\beta }
\def\Kab{K^{{\bf a}}_{\ {\bf b}}}
\def\hf{{1\over 2}}
\def\pl{\hbar}

\def\qgs{{\it quasi-ground-state}\/}
\def\KUNSmark{\vtop{\hbox{\elevenmib Department\hskip1mm of\hskip1mm
             Physics}\hbox{\elevenmib Kyoto\hskip1mm University}}}
\newif\ifKUNS \KUNStrue
\unlock
\def\titlepage{\FRONTPAGE\paperstyle\ifPhysRev\PH@SR@V\fi
    \ifKUNS\null\vskip-17mm\KUNSmark\vskip0mm\fi
    \ifp@bblock\p@bblock\fi}
\lock
\Pubnum={\ifdraft\undertext{\strut$\mib draft$}\cr\fi
         KUNS\the\pubnum}
\newcount\YEAR
\YEAR=\number\year
\global\advance\YEAR by-1900
\pubnum={1267}
\titlepage
\title{Quantum State During and After O(4)-Symmetric
Bubble Nucleation with Gravitational Effects}

\author{Takahiro~Tanaka and Misao~Sasaki}

\address{Department of Physics,~Kyoto University,
         ~Kyoto 606-01,~Japan}
\vskip1cm
\centerline{\it Submitted to Phys. Rev. D}
\abstract{
We extend our previous analysis of
the quantum state during and after $O(4)$-symmetric bubble nucleation
to the case including gravitational effects.
We find that there exists a simple relationship between the case
with and without gravitational effects. In a special case
of a conformally coupled scalar field which is massless except on
the bubble wall, the state is found to be conformally equivalent to
the case without gravity.
}
\REF\Inf{
  D.~La and P.~J.~Steinhardt, Phys. Rev. Lett. {\bf 62} (1989) 376;
\subpar
  P.~J.~Steinhardt and F.~S.~Accetta, Phys. Rev. Lett. {\bf 64}
  (1990) 2740;
\subpar
  F.~C.~Adams and K.~Freese, Phys. Rev. {\bf D43} (1991) 353;
\subpar
  R.~Holman, E.~W.~Kolb, S.~L.~Vadas and Y.~Wang,
  Phys. Lett. {\bf B269} (1991) 252;
\subpar
  A.~R.~Liddle and D.~Wands, Phys. Rev. {\bf D45} (1992) 2665.
}
\REF\SaTaYY{
  M. Sasaki, T. Tanaka, K. Yamamoto and J. Yokoyama,
 Phys. Lett. {\bf 317B} (1993) 510.
}
\REF\Ruba{
  V.A.~Rubakov, Nucl. Phys. {\bf B245} (1984) 481.
}
\REF\VacVil{
  T.~Vachaspati and A.~Vilenkin, Phys. Rev. {\bf D43} (1991) 3846.
}
\REF\TaSaYa{
 T. Tanaka, M. Sasaki and K. Yamamoto,
Phys. Rev. {\bf D49} (1994) 1039, [Paper I].
}
\REF\VBubble{
  M. Sasaki, T. Tanaka, K. Yamamoto and J. Yokoyama,
Prog. Theor. Phys. {\bf 90} (1993) 1019, [Paper II].
}
\REF\Yamamoto{
K.~Yamamoto, Prog. Theor. Phys. {\bf 91} (1994) 437.
}
\REF\Col{
  S. Coleman, V. Glaser and A. Martin, Commun. Math. Phys.
{\bf 58} (1978) 211;\subpar
  S. Coleman, in
  {\it The Whys of Subnuclear Physics,} ed.
    A. Zichichi (Plenum, New York, 1979), p.805.
}
\REF\ColDlu{
S. Coleman and F. De Luccia, Phys. Rev. {\bf D21} (1980) 3305.
}
\REF\TanSas{
  T. Tanaka and M. Sasaki,
 Prog. Theor. Phys. {\bf 88} (1992) 503.
}
\REF\Allen{
 B. Allen, Phys. Rev. {\bf D32} (1985) 3136.
}
\REF\BirDav{
W.~G.~Unruh Phys. Rev. {\bf D14} (1976) 870;\subpar
N.D. Birrell and P.C.W. Davies, {\it Quantum fields in curved space},
(Cambridge University Press, Cambridge, 1982).
}
\REF\Anomaly{
P. Candelas and J.S. Dowker, Phys. Rev. {\bf D19} (1979) 2902.
}

\chapter{Introduction}

{}Field theoretical quantum tunneling phenomena such as
false vacuum decay are considered to have played important
roles in the dynamics of the universe in its early stage.
One good example is the so-called extended inflation [\Inf],
in which the inflationary stage of the universe ends with
nucleation of true vacuum bubbles and thermalization of the universe
by collisions of these nucleated bubbles.

As another interesting possibility, we have recently proposed
a simple one-bubble scenario of the inflationary universe
by considering the particle creation
during and after the false vacuum decay [\SaTaYY].
Provided enough entropy is produced,
it is possible to have our universe inside one nucleated bubble.
However, since our knowledge of the quantum state
after the false vacuum decay is far from sufficient,
we are unable to argue further for or against this possibility
 at present.

Besides the inflationary universe scenario, what happens after
the bubble nucleation is an interesting issue as a
fundamental process relating to the quantum matter production or
the quantification of quantum effects in general
in the early universe.

Toward a clear understanding of the issue,
various attempts have been made.
Among them, Rubakov developed a method of non-unitary Bogoliubov
transformation to treat the particle production during tunneling
[\Ruba]. Then Vachaspati and Vilenkin investigated general
features of the quantum state during and after nucleation of
an $O(4)$-symmetric bubble, paying particular attention to the
symmetry of the state [\VacVil].
Meanwhile, we have developed a method to analyze the
quantum state during and
after field-theoretical quantum tunneling by constructing
a multi-dimensional wave function in a covariant manner
[\TaSaYa] (hereafter Paper I).
Then we have applied it to the $O(4)$-symmetric bubble nucleation
and investigated the properties of the quantum state of fluctuating
degrees of freedom in detail by
 constructing an analytically soluble model
[\VBubble] (hereafter Paper II).

However, all of these previous analyses were based on several
non-trivial assumptions or simplification of models,
the validity of which is not clear.
To mention one of such,
the effect of gravity was neglected in all of them.
In this paper, we focus on this point and tackle the problem to
incorporate the effect of gravity.
More precisely, as a first step, we take into account the
background spacetime curvature induced by the
tunneling field solution
and investigate its effect on the quantum state of fluctuating
degrees of freedom. Hence, in particular, the false vacuum is
de Sitter space.

This paper is organized as follows.
In section 2, we extend our method developed in Paper I to
the case when the initial state is excited with respect to
the false vacuum, which is necessary to incorporate the gravitational
effect.
In section 3, we show that there is an elegant interrelation between
the quantum states after tunneling with and without
the effect of gravity. As a specific example, we then consider
a conformally coupled scalar field which interacts with the tunneling
field on the bubble wall but massless elsewhere, and
show that its quantum state is conformally equivalent to the case
without gravity, i.e., the same result as for the Minkowski background
obtained in Paper II.
However, we also point out a paradoxical situation we encounter
when evaluating the regularized energy momentum tensor.
In section 4, we summarize our results.

\chapter{Multi-dimensional Tunneling Wave Function}
In Paper I, we developed a method to construct the
multi-dimensional tunneling wave function from the false vacuum ground
state in a covariant manner.
As we shall see in the next section, in order to construct the tunneling
wave function in curved spacetime, it is necessary to extend the
formalism given in Paper I to the case when the initial state is
in an excited state. Some basic parts of this extension have been
recently given by Yamamoto [\Yamamoto].

We consider a system of $D+1$ degrees of freedom whose Lagrangian
is given by
$$
\eqalign{
 L=\hf\Gab(\phi)\dot{\phi}^\alpha\dot{\phi}^\beta&-V(\phi),\cr
    & (\alpha,\beta=0,1,\cdots,D;~i,j=1,\cdots,D),\cr
}\eqn\tLag
$$
 where $\phi^\alpha$ are the coordinates for the $D+1$-dimensional
space of dynamical variables (\ie, superspace)
and $\Gab$ is the superspace metric.
 In this section, Greek and Latin indices run from $0$ to $D$
and from $1$ to $D$, respectively. For simplicity,
 we assume the potential $V(\phi)$ of the form,
$$
V(\phi)=U(X)+\hf m^2_{ij}(X)\phi^i\phi^j\,,
\ee
$$
where the tunneling degree of freedom is represented by
$X=\phi^0$ as a collective coordinate and the fluctuating
degrees of freedom by $\phi^i$, respectively. Further
we focus on the case when the superspace metric
depends only on the tunneling degree of freedom, $\Gab=\Gab(X)$,
and has no cross-terms between $X$ and $\phi^i$ (\ie,  $G_{0i}=0$),
and assume that the signature of the metric is
 positive definite.
 The potential $U(X)$ is supposed to have a local minimum
 at $X=X_F$, which is not the absolute minimum, as shown in Fig.~1.
 We call the point $(X,\phi^i)=(X_F,0)$ the false vacuum origin
throughout this paper.

The Hamiltonian operator in the coordinate representation
is obtained by replacing the conjugate
 momenta in the Hamiltonian with the corresponding
 differential operators.
 In general, there exists the operator ordering ambiguity.
Here we fix it in such
a way that the resulting Hamiltonian takes the form,
$$
 \hat H=-{\pl^2\over 2}\Gabu(X)\nba\nbb+U(X)
 +\hf m^2_{ij}(X)\phi^i\phi^j,
\eqn\tHam
$$
 where $\Gabu(X)$ is the inverse matrix of $\Gab(X)$.

 In Paper I, we constructed the \qgs wave function using
 the WKB approximation.
The \qgs wave function is the lowest eigenstate of the Hamiltonian
sufficiently localized at the false vacuum.
Let us briefly summarize the method without rigor.
The detailed discussion is given in Paper I.

\noindent(i)  First, we impose the WKB ansatz on the wave function,
$$
 \Psi=e^{-{1\over \pl}(W^{(0)}+\pl W^{(1)}+\cdots)},
\eqn\tPsi
$$
 which should solve the time-independent Schr{\"o}dinger equation,
$$
 \hat H\Psi=E\Psi.
\eqn\tBas
$$
 We solve this equation  order by order with respect to $\pl$.
 The energy eigenvalue $E$ is formally divided into two parts,
 $E_0$ and $E_1$, of $O(\pl^0)$ and $O(\pl^1)$, respectively.
 The equation in the lowest order of $\pl$ becomes the
 Hamilton-Jacobi equation with the energy $E_0\,$,
$$
  -\hf\Gabu\nba W^{(0)}\nbb W^{(0)}+V(\phi)=E_0.
\eqn\tlow
$$
 By setting $\Gabu\nbb W^{(0)}=\dot\phi^\alpha$, Eq.\tlow\ gives the
Euclidean equation of motion, \ie, with respect to the imaginary
time $\tau=it$.

\noindent(ii) We consider a solution of the Euclidean
equation of motion which starts from the false vacuum at
$\tau=-\infty$ with the zero kinetic energy (\ie, $E=E_0:=U(X_F)$)
and arrives at the turning point at $\tau=0$ which is
 the boundary of the classically allowed and forbidden regions.
If there are several nontrivial solutions, we choose the one which
gives the minimum Euclidean action.
we call it the dominant escape path (DEP), and denote it by
$\phi^\alpha_{(0)}(\tau)$.
In the present case, we have
$$
\left(\phi^0_{(0)}(\tau),\phi^i_{(0)}(\tau)\right)
 =\left(X(\tau),0\right).
\ee
$$

\noindent(iii) Next, along DEP, we introduce an orthonormal basis,
$\hat e^i_{\bf a}(\tau)$, lying in the hypersurface
$\tilde\Sigma(\tau)$ orthogonal to it;
$G_{ij}(X(\tau)) \hat e^i_{\bf a}(\tau)\hat e^j_{\bf b}(\tau)
=\delta_{\bf ab}$,
where ${\bf a}$ runs through the range $1,2, \cdots,D$. For
 convenience, we fix the orthonormal basis
in such a way that $e^i_{\bf a}:=\hat e^i_{\bf a}(\tau= -\infty)$
diagonalizes $\omega_{\bf ab}:=\omega_{ij}e^i_{\bf a}e^j_{\bf b}$,
\ie, $\omega_{\bf ab}=\omega_{\bf a}\delta_{\bf ab}$, where
$$
 \omega^2_{ij}:=\lim_{X\rightarrow X_F} m^2_{ij}(X).
\ee
$$
Then the required orthonormal basis along DEP is constructed
by solving the equation,
$$
 {\partial\over\partial\tau}\hat e^i_{\bf a}
 +\hf G^{ik} \dot G_{kj} \hat e^j_{\bf a}=0,
\ee
$$
which corresponds to a special case of Eq.(2.17) in Paper I.

\noindent(iv) Assuming that $W^{(0)}(\phi)$ is known around DEP,
we span the hypersurface orthogonal to DEP by the coordinates
$\eta^{\bf a}$ with respect to the basis $e^i_{\bf a}$ and
define
$$
 \tilde\Omega_{\bf ab}:=\left.W^{(0)}_{;ij}\right|_{\eta^{\bf c}=0}
      \hat e^i_{\bf a} \hat e^j_{\bf b},
\ee
$$
where the semicolon represents the $(D+1)$-dimensional covariant
differentiation with respect to $G_{\alpha\beta}$.
We also introduce a matrix $\Kab$ which is determined by solving
$$
\eqalign{
 K^{\bf a}_{\bf b} & =\hat e^{\bf a}_\alpha K^\alpha_{\bf b},\cr
  {D^2\over d\tau^2}K^\alpha_{\bf b} & =
  \left(\left(U+\hf m^2_{ij}\phi^i\phi^j\right)
 ^{;\alpha}_{~;\beta}-\dot X(\tau)^2
 R^\alpha_{~0\beta0} \right)K^\beta_{\bf b},\cr
}\eqn\Keqb
$$
with the boundary condition,
$$
  \Kab\rightarrow\bigl(e^{{\omega}\tau}\bigr)^{\bf a}_{~\bf b},
  ~~~~(\tau\rightarrow -\infty),
\eqn\Kasymp
$$
where $D/d\tau$ and $R^\alpha_{\gamma\beta\rho}$ denote
the Lie derivative along DEP and the Riemann tensor of
the superspace metric $G_{\alpha\beta}$, respectively.
One then finds $\tilde\Omega_{\bf ab}$ is expressed in terms of
$K_{\bf ab}$ as
$$
  \tilde\Omega_{\bf{ab}}=
  {\dot K_{\bf a}^{~\bf c} } \bigl({K^{-1}}\bigr)_{\bf c\bf b}.
\eqn\azzkk
$$

\noindent(v) With these results in hand,
the \qgs wave function [\TaSaYa] is found to be
$$
\eqalign{
  \Psi=&{C\Bigl( \det \omega/\pi\pl \Bigr)^{1/4}e^{\hf\omega_0\tau}
 \over
  \Bigl[2\bigl( U(X(\tau))-E_0 \bigr)\Bigr]^{1/4}
  \sqrt{ \big\vert \det\Kab(\tau) \big\vert} }
\cr
  &\times\exp\Bigl(-{1\over\pl}\int_{-\infty}^{\tau}d\tau'
  2\bigl(U(X(\tau))-E_0 \bigr)
 +{1\over2}\Tr\omega\tau
  -{1\over2\pl}\tilde\Omega_{\bf{ab}}\eta^{\bf a}\eta^{\bf b}\Bigr),
\cr
}
\eqn\fina
$$
where $\omega_0:=U''(X_F)$, $\Tr\omega=\sum_{{\bf a}=1}^D\omega_{\bf a}$
, $\det\omega=\prod_{{\bf a}=1}^D\omega_{\bf a}$
and the normalization constant $C$ is given by
$$
 C=\lim_{\tau\rightarrow -\infty}
    {\left(2\bigl(U(X(\tau))-E_0 \bigr)\right)^{1/4}
    \over e^{\hf\omega_0 \tau}}
\left({\omega_0\over\pi\hbar}\right)^{1/4}.
\eqn\matchb
$$

\bigskip

In the present case, the above result can be re-expressed in terms
of the superspace coordinates $(\tau,\phi^i)$ as follows. First,
 note that the hypersurface $\tilde\Sigma(\tau)$ is generally
different from the $\tau=const.$ hypersurface
$\Sigma(\tau)$ off the DEP; the latter is warped if $\dot G_{ij}\neq0$.
Let a point $(\tilde\tau,\eta^{\bf a})$ on the hypersurface
$\tilde\Sigma(\tilde\tau)$ correspond to $(\tau,\phi^i)$ in the
original coordinates.
Then from the fact $\tilde Q(\tilde\tau,\eta^{\bf a})=Q(\tau,\phi^i)$
for any scalar function $Q$ and the equation,
$$
 \tilde\Omega_{ij}=\Omega_{ij}+\hf\dot G_{ij},
\eqn\OmeOme
$$
which follows from the definition of the covariant derivative,
where
$$
 \Omega_{ij}:=\left.
          {\partial^2\over\partial\phi^i\partial\phi^j}W^{(0)}
              \right|_{\phi^k=0},
\quad
 \tilde\Omega_{ij}:=\hat e^{\bf a}_i\hat e^{\bf b}_j\tilde\Omega_{\bf ab}\,,
\ee
$$
we find
$$
 \tilde\tau=\tau-\hf \left({dW^{(0)}\over d\tau}\right)^{-1}
         \dot G_{ij}\phi^i\phi^j\,.
\eqn\tautau
$$

Now, replacing $\tau$ in Eq.\fina\ with $\tilde\tau$ given by Eq.\tautau,
expanding the result around $\tau$,
noting the fact $dW^{(0)}/d\tau=2(U(X)-E_0)$, and regarding $\eta^{\bf a}$
and $\phi^i$ as quantities of $O(\hbar^{1/2})$, we find the wave
function to the first WKB order to have the form,
$$
 \Psi(\phi^\alpha)=\Theta(X) \Phi(X,\phi^i),
 \eqn\Eqdec
$$
where $\Theta$ is the lowest WKB part,
$$
  \Theta(X)={C \over  \Bigl[2\bigl( V(X(\tau))-E_0 \bigr)\Bigr]^{1/4}}
  \exp\Bigl(-{1\over\pl}\int_{-\infty}^{\tau}d\tau'
  2\bigl(U(X(\tau))-E_0\bigr)+{1\over 2}\omega_0 \tau \Bigr),
\eqn\eqtheta
$$
and $\Phi$ is the first WKB correction,
$$
  \Phi(X,\phi^i)={\Bigl( \det \omega/\pi \Bigr)^{1/4}
  \over \sqrt{ \big\vert \det K^i_{\bf a}(\tau)\sqrt{G}\big\vert} }
  \exp\Bigl({1\over2}\Tr\omega\tau-{1\over2\pl}
        \Omega_{ij}\phi^i\phi^j\Bigr).
\eqn\finaff
$$

{}Further, from Eqs.\azzkk\ and \OmeOme, we obtain
$$
  \Omega_{ij}=\sum_{\bf a}G_{ik}\dot K^k_{\bf a} K^{-1}_{j\bf a}.
\ee
$$
Also, from Eq.\Keqb, we find $K^i_{\bf a}$ satisfies the equation,
$$
  G_{ij}\ddot K^j_{\bf a}+\dot G_{ij}\dot K^j_{\bf a}=
  m^2_{ij}K^j_{\bf a}\,,
\eqn\Keq
$$
which is just the classical equation of motion for $\phi^i$
in the Euclidean time.
With the help of Eq.\Keq, it is then straightforward
to show that to the first WKB order \finaff\ satisfies
$$
\eqalign{
\left[{\cal D}-E'_1\right]&\Phi=0\,;
\cr
&{\cal D}:= \left[G^{-1/4}\partial_\tau G^{1/4}
   -{\pl\over 2}G^{ij}\partial_i\partial_j
   +{1\over 2\pl}m^2_{ij}\phi^i\phi^j\right],
\cr}
\eqn\eqphi
$$
where $E'_1=\hbar\Tr\omega$. One sees that
this is just the Euclidean time Schr\"odinger equation
for the fluctuating degrees of freedom.

{}Following the procedure taken in [\Yamamoto], we now construct
a set of generalized creation and annihilation operators,
$A^{\dag}_{\bf a}$ and $A_{\bf a}$ whose action on some eigenstate of
the Hamiltonian produces another eigenstate. In other words, look for
operators which correspond to the usual creation and
annihilation operators at the false vacuum origin, \ie,
$[{\cal D},A_{\bf a}]=\omega_{\bf a}A_{\bf a}$ and
$[{\cal D},A^{\dag}_{\bf a}]=-\omega_{\bf a}A^{\dag}_{\bf a}$.
To do so, first, we make a set of operators $a_{\bf a}$ and
$a^{\dag}_{\bf a}$ which
commute with the differential operator ${\cal D}$.
If we assume the forms of $a_{\bf a}$ and $a^{\dag}_{\bf a}$ as
$$
\eqalign{
 \hbar a_{\bf a}= &\sqrt{\pl\over 2\omega_{\bf a}}K^i_{\bf a}
         \hbar{\partial\over\partial\phi^i}
      +G_{ij}\sqrt{\pl\over 2\omega_{\bf a}}\dot K^i_{\bf a}
         \phi^j, \cr
 \hbar a^{\dag}_{\bf a}= &
       -\sqrt{\pl\over 2\omega_{\bf a}}Q^i_{\bf a}
       \hbar{\partial\over\partial\phi^i}
      -G_{ij}\sqrt{\pl\over 2\omega_{\bf a}}\dot Q^i_{\bf a}
       \phi^j, \cr
}\ee
$$
it is easy to see that $a_{\bf a}$ commute with $\cal D$ and
the necessary condition that $a^{\dag}_{\bf a}$ commute with
$\cal D$ is that $Q^i_{\bf a}$ satisfies the same equation as Eq.\Keq\
for $K^i_{\bf a}$.
Then, as we have adopted the orthonormal basis $\hat e^i_{\bf a}$
which diagonalizes $\omega_{\bf ab}$ at $\tau\rightarrow-\infty$,
if we set the boundary condition of $Q^i_{\bf a}$ as
$$
 Q^i_{\bf a} \rightarrow
      e^{-\omega_{\bf a}\tau} e^i_{\bf a}\quad{\rm for}~
         \tau\rightarrow-\infty\,,
\eqn\InitK
$$
we find $[a^{\dag}_{\bf a},a_{\bf b}]=\delta_{\bf ab}$.
Hence the relevant creation and annihilation operators
 $A^{\dag}_{\bf a}$ and $A_{\bf a}$ are found to be
$$
\eqalign{
 A_{\bf a} & =e^{-\omega_{\bf a}\tau}a_{\bf a},
 \cr
 A^{\dag}_{\bf a} & =e^{\omega_{\bf a}\tau}a^{\dag}_{\bf a}.
 \cr
}\ee
$$
An excited state wave function with respect to
the fluctuating degrees of freedom can be obtained by operating these
creation operators, $A^{\dag}_{\bf a}$, to the \qgs wave function
\finaff.

Given these results, it is convenient to reformulate the method
to construct the wave function in the following way.
We consider a path in the complex plane of time as shown
in Fig.~2. Along this path, we construct a solution $X(t)$ of the
classical equation of motion with the initial condition
$X(-\infty)=X_F$.
In the segment $A$ of the path, the solution stays at
the false vacuum origin $X(t)=X_F$ which is certainly a solution.
This solution can be smoothly connected to the DEP solution at
 a sufficient large negative $\tau$ ($=it$).
The segment $B$ corresponds to the DEP in the forbidden region
and $C$ the allowed region. The solution along $C$ is obtained by
analytically continuing the DEP solution by setting $t=-i\tau$ ($>0$),
which describes the classical motion after tunneling.
The lowest WKB order wave function is described by
this classical solution.
Now along this path, we solve Eq.\Keq\ for
$K^i_{\bf a}$ and $Q^i_{\bf a}$
with the initial conditions at $t\rightarrow-\infty$ as
$$
 \eqalign{
 \sqrt{\pl\over 2\omega_{\bf a}}Q^i_{\bf a} & =u^i_{\bf a},\cr
 \sqrt{\pl\over 2\omega_{\bf a}}K^i_{\bf a} & =u^{i*}_{\bf a},\cr
}
\ee
$$
where $u^i_{\bf a}$ and $u^{i*}_{\bf a}$ are the positive and
negative frequency functions, respectively, in the false vacuum.
In this way, we obtain the first WKB order wave function before and
after the tunneling.

To close this section, perhaps it is worthwhile to mention that
the time $t$ (or $\tau$) discussed here is {\it not\/} the external
 time in the original Schr\"odinger equation but is a parameter that
naturally arises from the characterization of the lowest
WKB configuration (or the internal time).

\chapter{Incorporation of Gravitational Effect}

We consider the system which consists of
two scalar fields, \ie, the tunneling field $\sigma$ and another field
$\phi$ which represents the fluctuating degrees of
freedom, both coupled to gravity.
We consider the situation in which the potential $U(\sigma)$
is in the form shown in Fig.~1 and $\sigma$ is initially
at the false vacuum, $\sigma=\sigma_F$.
The interaction between the two fields
is assumed to be described by the $\sigma$-dependent mass
term of $\phi$; $m^2(\sigma)\phi^2/2$.
As in Paper II, we ignore the fluctuations of $\sigma$ and the metric
$g_{\mu\nu}$ for simplicity.
We begin with the Lagrangian of the form,
$$
 {\cal L}={\cal L}_{grav}+{\cal L}_\sigma+{\cal L}_\phi,
\eqn\tLag
$$
 where
$$
 \eqalign{
 {\cal L}_{grav} & =\alpha\sqrt{\gamma}{1\over 16\pi G}R,\cr
 {\cal L}_\sigma & =-\alpha\sqrt{\gamma}\left[
      \hf g^{\mu\nu}\nabla_{\mu}\sigma\nabla_{\nu}\sigma
       +U(\sigma)\right],\cr
 {\cal L}_\phi & =-\alpha\sqrt{\gamma}\left[
      \hf{g}^{\mu\nu}\nabla_{\mu}\phi\nabla_{\nu}\phi
       +\left(m^2(\sigma)+\xi R\right)\phi^2\right],\cr
}
\eqn\eLag
$$
and $G$ and $R$ are the gravitational constant and the scalar
curvature, respectively.
Here and in what follows, we use the notation of the $3+1$
decomposition of the spacetime metric:
$$
 {g}_{\mu\nu}=\left(\matrix
       {-\alpha^2+\beta_s \beta^s & \beta_n\cr
        \beta_m & \gamma_{nm}\cr}\right).
\ee
$$
The Hamiltonian is given by
$$
 H=H_{grav}+H_\sigma+H_\phi,
\eqn\hhami
$$
 where $H_{grav}$ and $H_\sigma$ are the Hamiltonians of
the gravitational and tunneling fields, respectively, and
$H_\phi$ is that of the fluctuating field,
$$
 H_\phi =\int d^3x \Bigl[~\hf{\alpha\over\sqrt{\gamma}} p^2
         +p\beta^m\nabla_m\phi
         +{\alpha\sqrt{\gamma}\over 2}
         \left(\gamma^{km}  \nabla_k\phi\nabla_m\phi+\left(m^2(\sigma)
         +\xi R\right)\phi^2\right)\Bigr],
\eqn\hHel
$$
with $p$ being the momentum conjugate to $\phi$.

We should note that when gravity comes into play, there exists no
external time and the Schr\"odinger equation becomes the Wheeler-DeWitt
equation,
$$
 H\Psi=0,
\ee
$$
which is essentially the Hamiltonian constraint for the total system
and the superspace metric has an indefinite signature, presumably with
only one timelike component. However, discussion of the nature
of the Wheeler-DeWitt equation is out of the scope of this paper.
Since we ignore fluctuations in $g_{\mu\nu}$ as well as in $\sigma$,
we simply assume that the lowest WKB state is described by a classical
solution of the $(\sigma,g_{\mu\nu})$ system and ignore problems associated
with the Wheeler-DeWitt equation. Then, as we shall see below,
there arises no conceptual problem with the construction of the wave
functional for $\phi$.

\section{Instanton with gravity}

Let us first construct a non-trivial solution of the Euclidean
Einstein-scalar field equations, an instanton (or bounce) with gravity,
 to obtain the lowest WKB order picture.
In the absence of gravity, it has been shown that the classical
solution with the minimum action is $O(4)$-symmetric [\Col].
 Although it is not proved when gravity is present,
it seems reasonable that it is also
the case in the presence of gravity. Hence we assume so.
The $O(4)$-symmetric instanton with gravity was
investigated by Coleman and De Luccia [\ColDlu]. Here we shall
not repeat the details but discuss only those features of the instanton
that will be necessary for our purpose.

The $O(4)$-symmetric instanton takes the form,
$$
 \eqalign{
 ds_E^2 & =n^2(\eta)d\eta^2+a^2(\eta)
            \left(dr^2+\sin^2rd\Omega^2_{(2)}\right),\cr
 \sigma& =\sigma(\eta),\cr
 \phi& =0.\cr
 }
\eqn\metric
$$
Since we ignore the fluctuations in $\sigma$ and $g_{\mu\nu}$,
we denote the instanton configuration simply by $\sigma(\eta)$ and
 $g_{\mu\nu}(\eta)$.
Then the Euclidean action becomes
$$
 S_E=2\pi^2\int d\eta\left[
      {1\over n}\left(
        -{3\over 8\pi G}a\dot a^2+\hf a^3\dot\sigma^2\right)
    +n\left(-{3a\over 8\pi G}+a^3 U(\sigma)\right)\right],
\eqn\redact
$$
where a dot means the derivative with respect to $\eta$.
We see that $n$ plays the role of a Lagrange multiplier;
a consequence of the time reparametrization invariance of the system.
The variation of $S_E$ with respect to $\sigma$ gives
$$
   \ddot\sigma+\left(3{\dot a\over a}-{\dot n\over n}
   \right)\dot\sigma={dU\over d\sigma},
\eqn\feqs
$$
 and that with respect to $n$ gives
$$
 \dot a^2-{4\pi G\over 3}a^2\dot\sigma^2=n^2
   \left(1-{8\pi G\over 3}a^2 U(\sigma)\right),
\eqn\constraint
$$
which is nothing but the Hamiltonian constraint.
Because of this constraint, the variation with respect to $a$
does not give an independent equation.

Let us present the solution of the above equations in the thin-wall
case. We also assume that
the true vacuum energy density $U(\sigma_T)$ is non-negative.
We choose the gauge $n(\eta)=a(\eta)$ and indicate
the wall position by $\eta=\eta_W$. The result is
$$
\eqalign{
&a(\eta)=\cr
&\cases{\displaystyle{1\over H_F \cosh\eta}\qquad(\eta<\eta_W),&\cr
          \displaystyle{1\over H_F\left(
                 \cosh(\eta-\eta_W)\cosh\eta_W+
        \sinh(\eta-\eta_W)\sqrt{\cosh^2\eta_W-(H_T/H_F)^2}
                                  \right)}&\cr
       \phantom{\displaystyle{1\over H_F \cosh\eta}\qquad}
                           (\eta>\eta_W),&\cr}
\crr
&\sigma(\eta)=
 \cases{\sigma_F\quad &($\eta<\eta_W$),\cr
         \sigma_T\quad &($\eta>\eta_W$),\cr}
\cr}
\eqn\thwall
$$
where $\eta_W>0$ always and
$$
H_F^2={8\pi G\over 3}U(\sigma_F),
\quad H_T^2={8\pi G\over 3}U(\sigma_T).
\ee
$$
A schematic picture of the instanton is shown in Fig.~3.
It has the topology of $S^4$.
In the thin-wall approximation, the metric and tunneling field
configurations are identical to those of the false vacuum
at $\eta<\eta_W$, hence at $\eta<0$.
Here, we restrict our attention to the case this holds,
but it should be mentioned that this is not true in general once
the thin-wall approximation breaks down.

As was pointed out in [\TanSas], the coordinate $\eta$
cannot play the role of the `time' parameter $\tau$ which
distinguishes each spatial configuration of the instanton
the sequence of which connects the false vacuum and the critical bubble
configuration corresponding to the turning point.
A relevant choice of it is obtained by
the following coordinate transformation,
$$
\left\{
\eqalign{
 &\cosh\eta ={1\over\sqrt{\sin^2\tau+R^2\cos^2\tau}}\,,
\cr
 &\sin r ={R\over\sqrt{\sin^2\tau+R^2\cos^2\tau}}\,,
\crr
&\qquad(-\infty<\eta<\infty,~0\leq r\leq\pi),
\cr}\right.
{}~\Leftrightarrow
\left\{
\eqalign{
 &\sin\tau=-{\cos r\over\sqrt{\cosh^2\eta-\sin^2r}}\,,
\cr
 &R={\sin r\over\cosh\eta}\,,
\crr
&\qquad(-\pi\leq\tau\leq\pi,~0\leq R\leq1).
\cr}\right.
 \eqn\ctrans
$$
Then we have
$$
 ds_E^2=\cosh^2\eta~a^2(\eta)\left\{
    (1-R^2)d\tau^2+{dR^2\over 1-R^2}+R^2d\Omega^2_{(2)}
    \right\}.
 \eqn\static
$$
How these coordinates span the Euclidean spacetime is schematically
shown in Fig.~4.

There are two main differences from the original coordinates. First,
 these coordinates reduce to the static de Sitter coordinates
when $a(\eta)=1/H_F \cosh\eta$.
Therefore at $\tau<-\pi/2$,
the spatial metric and the tunneling field configuration
on each $\tau=const.$ surface are identical to those at
the false vacuum origin
and it is also possible to extend this solution beyond $\tau=-\pi$
to $\tau=-\infty$.
Second, the configuration on the $\tau= 0$ surface corresponds
to the turning point, where the solution can be
analytically continued to the Lorentzian
region by $t=-i\tau$ ($>0$).
How these coordinates are analytically continued to the
Lorentzian region is shown in Fig.~5.
Using these coordinates, a full description of
the false vacuum decay at the lowest WKB order is obtained and
our formalism of the multi-dimensional wave function can
be applied to investigate the next WKB order effects.

\section{Quantum state of the fluctuating field}

Next we turn to the issue of the quantum fluctuations
during and after this quantum tunneling.
As mentioned previously, we neglect the fluctuations of the
metric and the tunneling field to avoid difficulties. For
convenience, we use the symbol $X$ to represent
the tunneling degree of freedom in the superspace, \ie,
the spatial configurations of the metric and the tunneling field,
$g_{\mu\nu}=g_{\mu\nu}(X)$ and $\sigma=\sigma(X)$.
We assume $X$ is suitably normalized so that the
reduced Hamiltonian takes the form,
$$
 H_{red}=-{\pl^2\over 2}{\partial^2\over\partial X^2}+U(X)
       +H_\phi(X,\phi).
\eqn\hred
$$
We then investigate the quantum state described by $H_\phi(X,\phi)$
with $X$ now representing the DEP parametrized by $\tau$, $X=X(\tau)$.

To apply our formalism developed in the previous section to the present
case, we replace the suffices $i,j,\cdots$ with the spatial
coordinates ${\mib x,y,}\cdots$ and
 ${\bf a,b,}\cdots$ with certain eigenvalue
indices ${\mib k,p,}\cdots$ for a complete set of mode functions,
say $w_{\mib k}$:
$$
\eqalign{
   & \phi^i\rightarrow \phi\left({\mib x}\right),
\qquad
  \sqrt{\pl\over 2\omega_{\bf a}}
  K^i_{\bf a}(\tau)\rightarrow  w_{\mib k}\left({\mib x},\tau\right),
\cr
   & G^{ij}(\tau)\rightarrow
  {\alpha\left({\mib x},\tau\right)\over
    \sqrt{{\gamma}\left({\mib x},\tau\right)}}
     \delta^3\left({\mib x}-{\mib y}\right).\cr
}\eqn\corre
$$
Here one comment is in order. If we wish to deal with
the spacetime metric with non-vanishing shift vector $\beta^i$,
we would need to generalize the formalism in \S 2 to
allow the superspace metric to have $G_{0i}\neq0$.
However, since there is no shift vector in our metric, Eq.\static,
this generalization is unnecessary at the moment.

Now as the state before tunneling, the most natural false vacuum
state is the so-called Euclidean vacuum, which is de Sitter invariant
and exhibits the same short distance behavior of the field
as the Minkowski vacuum [\Allen].
Although the latter property is essential to
single out the Euclidean vacuum, it can be made explicit only when
we deal with a specific theory.
Hence we focus on the de Sitter invariance of the vacuum in this
subsection.
Since the natural mode functions associated with the form of
the metric \static\ (with $a(\eta)=(H_F\cosh\eta)^{-1}$) does not
 respect the de Sitter invariance,
the relation between these mode functions and those for the Euclidean
vacuum is non-trivial.
Specifically, they are related by a Bogoliubov transformation.
This implies that the Euclidean vacuum is described as some kind of
an excited state relative to the ground state constructed with
respect to the static time coordinate
(we call the latter the static vacuum for convenience).
This is the reason why it was necessary to extend our formalism
to the case of excited states at false vacuum.

To prepare the Euclidean vacuum and to obtain the
Schr\"odinger wave functional relevant for the false vacuum decay,
in what follows, we first consider a general static spacetime and
overview the relation between
the usual Heisenberg representation of a vacuum state
and the corresponding Schr\"odinger wave functional.
Then we construct the Euclidean vacuum over the static vacuum and
translate the result to the Schr\"odinger picture.
Once this is done, it is then straightforward to obtain the tunneling
wave functional according to the prescription given in
the previous section.

Let $u_{\mib k}({\mib x},t)$ be a set of mode functions (not necessarily
the positive frequency functions with respect to
the static time coordinate)
and $A_{\mib k}$ be the corresponding
annihilation operator; $A_{\mib k}|O\rangle=0$, where $|O\rangle$ is
the ``vacuum" in the Heisenberg picture.
Then we have
$$
\eqalign{
 \hat\phi({\mib x},t)&
=\sum_{\mib k}\left( u_{\mib k}({\mib x},t) A_{\mib k}
             +u_{\mib k}^{*}({\mib x},t) A^{\dag}_{\mib k}\right),
\cr
 \hat p({\mib x},t)&
={\sqrt{\gamma({\mib x})}\over\alpha({\mib x})}
    \sum_{\mib k} \left(\dot u_{\mib k}({\mib x},t) A_{\mib k}
            +\dot u_{\mib k}^{*}({\mib x},t) A^{\dag}_{\mib k}\right).
\cr
}\eqn\modeex
$$
We assume the mode functions $u_{\mib k}({\mib x},t)$ are
orthonormalized with respect to the Klein-Gordon inner product,
$$
(u_{\mib k},u_{\mib p}):= -i\int d^3 x {\sqrt{\gamma}\over\alpha}
  \left(u_{\mib k}\dot u^{*}_{\mib p}
      -\dot u_{\mib k} u^{*}_{\mib p}\right)=\delta_{\mib kp}.
\eqn\normKG
$$
To consider the Schr\"odinger representation,
we introduce time-dependent annihilation and creation operators
 $a_{\mib k}(t)$ and $a^{\dag}_{\mib k}(t)$, respectively, as
$$
 a_{\mib k}(t)=e^{-i\hat Ht}A_{\mib k}e^{i\hat Ht},
\quad
 a^{\dag}_{\mib k}(t)=e^{-i\hat Ht}A^{\dag}_{\mib k}e^{i\hat Ht},
\eqn\Schancr
$$
where $\hat H$ is the Hamiltonian operator.
The Schr\"odinger representations of the field operators
$\hat\phi_S({\mib x})$ and $\hat p_S({\mib x})$ are given by
$$
\eqalign{
 \hat\phi_S({\mib x})
  & = e^{-i\hat Ht}\hat\phi({\mib x},t)e^{i\hat Ht}
\cr
  & =\sum_{\mib k} \left(u_{\mib k}({\mib x},t) a_{\mib k}(t)
             +u_{\mib k}^{*}({\mib x},t) a^{\dag}_{\mib k}(t)\right),
\cr
 \hat p_S({\mib x})
  & = e^{-i\hat Ht}\hat p({\mib x},t)e^{i\hat Ht}
\cr
  & ={\sqrt{\gamma({\mib x})}\over\alpha({\mib x})}
      \sum_{\mib k}\left( \dot u_{\mib k}({\mib x},t) a_{\mib k}(t)
       +\dot u_{\mib k}^{*}({\mib x},t) a^{\dag}_{\mib k}(t)\right).
\cr
}\eqn\tdadef
$$
Using these operators, the Schr\"odinger representation of
the vacuum, \ie, $\vert O(t)\rangle_S=e^{-i\hat Ht}\vert O\rangle$
 is determined by the condition,
$$
 a_{\mib k}(t)\vert O(t)\rangle_S=0.
 \eqn\tdvacdef
$$
On the other hand, using the orthonormality of the mode functions,
$a_{\mib k}(t)$ and $a_{\mib k}^{\dag}(t)$ are expressed as
$$
\eqalign{
a_{\mib k}(t) &
=i\int d^3x\left( u^{*}_{\mib k}({\mib x},t)\hat p_S({\mib x})
   -{\sqrt{\gamma}\over\alpha}
  \dot u^{*}_{\mib k}({\mib x},t)\hat\phi_S({\mib x})\right),
\cr
 a^{\dag}_{\mib k}(t) &
=i\int d^3x\left(- u_{\mib k}({\mib x},t)\hat p_S({\mib x})
   +{\sqrt{\gamma}\over\alpha}
  \dot u_{\mib k}({\mib x},t)\hat\phi_S({\mib x})\right).\cr
}\eqn\ancrexp
$$
Then, going over to the coordinate representation by
the replacements,
$$
\hat p_S({\mib x})\rightarrow -i\pl {\delta\over\delta\phi({\mib x})}\,,
\quad
\hat \phi_S({\mib x})\rightarrow\phi({\mib x})\,,
\ee
$$
we find from Eq.\tdvacdef\ that
$$
  \vert O(t)\rangle_S={\cal N}\exp\left(-{1\over 2\pl}
     \int\int d^3xd^3y~\Omega({\mib x},{\mib y};t)
            \phi({\mib x})\phi({\mib y})\right),
\eqn\Schrvac
$$
where ${\cal N}$ is a normalization constant and
$$
\Omega({\mib x},{\mib y};t)
 ={1\over i}{\sqrt{\gamma({\mib x})}\over \alpha({\mib x})}
      \sum_{\mib k}\dot u^*_{\mib k}({\mib x},t)
                        u^{*-1}_{\mib k}({\mib y},t),
\ee
$$
where $u^{-1}_{\mib k}$ is defined as
$$
 \sum_{\mib k}u_{\mib k}({\mib x},t)u^{-1}_{\mib k}({\mib y},t)
   =\delta^3({\mib x}-{\mib y}).
\ee
$$

Now let us specialize the above to the case of de Sitter space and
construct the Euclidean vacuum over the static vacuum. Since the
Euclidean vacuum is de Sitter invariant, we need a set of mode functions
which are defined over a complete Cauchy surface. However, it cannot be
covered by one static chart.  Just as in the case of describing the
Minkowski vacuum in terms of the Rindler mode functions in the Minkowski
spacetime [\BirDav], we therefore need to prepare two static charts.  We
label the quantities associated with these two charts by the indices (1)
and (2).  The two regions are causally disconnected. Furthermore, as
clear from the metric \static, the bubble nucleation takes place only in
one of the two regions. For convenience we regard the region (1) to be
the one in which the bubble nucleation occurs.  As the time direction is
opposite in the two regions, we fix it by identifying the future
direction with the time direction in the region (1).  Thus a complete
Cauchy surface is given by a hypersurface $t=t^{(1)}=t^{(2)}$ and the
Hamiltonian operator is positive in the region (1) and negative in the
region (2): $$ \hat H_{\phi}=\hat H_{\phi}^{(1)}-\hat H_{\phi}^{(2)},
\eqn\seph $$ where both the operators $H_{\phi}^{(i)}$ ($i=1,2$) are
positive and have the same form.

The positive frequency functions in each static chart
behave as
$$
 u^{(1)}_{{\mib k}}\propto e^{-i\omega_{\mib k}t^{(1)}},\quad
 u^{(2)}_{{\mib k}}\propto e^{i\omega_{\mib k}t^{(2)}},
 \eqn\stmode
$$
where $t^{(1)}$ and $t^{(2)}$ are the static time coordinates in the
regions (1) and (2), respectively.
We orthonormalize them by the Klein-Gordon inner product \normKG.
Note that the lapse function $\alpha$ is negative in the region (2).
The static vacuum is expressed as
$$
 |O\rangle=|O^{(1)}\rangle\otimes|O^{(2)}\rangle;
\quad
 A^{(i)}_{\mib k}|O^{(i)}\rangle=0\quad(i=1,2),
\eqn\statvac
$$
where $A^{(i)}_{\mib k}$ are
the annihilation operators associated with the positive frequency
functions $u^{(i)}_{{\mib k}}$.

The positive frequency functions for the Euclidean vacuum are
expressed in terms of a Bogoliubov transformation
from the static vacuum positive frequency functions.
Since the regions (1) and (2) are completely symmetric,
there exist two independent positive frequency functions
for the Euclidean vacuum for each $\mib k$,
which we denote by $\bar u^{(1)}_{\mib k}$ and
$\bar u^{(2)}_{\mib k}$.
Then in matrix notation with indices $\mib k$ suppressed,
the Bogoliubov transformation takes the form,
$$
\eqalign{
\left(\matrix{\bar u^{(1)}\cr\bar u^{(2)}\cr}\right)
&=\alpha\left(\matrix{u^{(1)}\cr u^{(2)}\cr}\right)
  +\beta\left(\matrix{u^{(1)*}\cr u^{(2)*}\cr}\right);
\cr
 \alpha&=\left(\matrix{\alpha^{(1)}&\alpha^{(2)}\cr
                       \alpha^{(2)}&\alpha^{(1)}\cr}\right),\quad
  \beta=\left(\matrix{\beta^{(1)}&\beta^{(2)}\cr
                      \beta^{(2)}&\beta^{(1)}\cr}\right),
\cr
 &\alpha\alpha^{\dag}-\beta\beta^{\dag}=I,
\cr}
\eqn\bog
$$
where $I$ denotes the unit matrix. Further,
by a unitary transformation, we can always
make the matrix $\alpha$ diagonal. Hence we may assume
$$
\alpha^{(1)}_{\mib kk'}=\alpha_{\mib k}\delta_{\mib kk'}\,,\quad
\alpha^{(2)}_{\mib kk'}=0.
\eqn\alal
$$

Let $|\bar O\rangle$ be the Euclidean vacuum. Then
it is characterized by
$$
 \bar A^{(i)}_{\mib k}|\bar O\rangle=0\quad (i=1,2),
\eqn\Eucldef
$$
where $\bar A^{(i)}_{\mib k}$ are
 the annihilation operator associated with the positive
frequency functions $\bar u^{(i)}_{\mib k}$.
The corresponding Bogoliubov transformation
 for $\bar A^{(i)}_{\mib k}$ is
$$
\left(\matrix{\bar A^{(1)}\cr\bar A^{(2)}\cr}\right)
   =\alpha^*\left(\matrix{A^{(1)}\cr A^{(2)}\cr}\right)
     -\beta^*\left(\matrix{A^{(1)\dag}\cr A^{(2)\dag}\cr}\right).
\eqn\Bog
$$
Then using the commutation relation between
creation and annihilation operators, we find that
the vacuum $|\bar O\rangle$ is expressed as
$$
 \vert\bar O\rangle={\cal N}
  \exp\left({\hf\sum_{i,j,{\mib k},{\mib k}'}
 B_{{\mib k}{\mib k}'}^{(i)(j)} A^{(i)\dag}_{\mib k}
      A^{(j)\dag}_{{\mib k}'}}\right)
 \vert O^{(1)}\rangle\otimes\vert O^{(2)}\rangle,
\ee
$$
where ${\cal N}$ is some normalization constant and
$$
 B_{{\mib k}{\mib k}'}^{(i)(j)}
  =\left(\alpha^{*-1}\beta^*\right)^{(i)(j)}_{{\mib k}{\mib k}'}.
\eqn\defB
$$

The Euclidean vacuum state is invariant
under the action of any generator of the de Sitter group.
Suppose the indices $\mib k$ of $u^{(i)}_{\mib k}$
represent the eigenvalues associated with spherical mode decomposition,
$(k,\ell,m)$, \ie, each of these modes is characterized by
an eigenstate of the Hamiltonian $\hat H_\phi$,
the angular momentum square
$\hat{\mib J}^2$, and the $z$-component of it $\hat J_z$.
Since these operators are generators of the de Sitter
group, the Euclidean vacuum must be a zero eigenstate
of all of them. From the fact that all of these operators
have the form, $\hat Q=\hat Q^{(1)}-\hat Q^{(2)}$,
we find the matrix $B$ should take the form,
$$
B=\left(\matrix{0&b\cr b&0\cr}\right);
\quad b_{\mib kk'}=B_{\mib k}\delta_{\mib kk'}\,.
\eqn\bogstate
$$
Then from Eqs.\alal, \defB\ and \bogstate,
the matrix $\beta$ is found to have the form,
$$
 \beta^{(1)}_{\mib kk'}=0,\quad\beta^{(2)}_{\mib kk'}
=\beta_{\mib k}\delta_{\mib kk'}\,,
\ee
$$
and
$$
 B_{\mib k}={\beta^*_{\mib k}\over\alpha^*_{\mib k}}\,.
\eqn\Bkdef
$$
Specifically, the Euclidean vacuum positive frequency functions
are given by
$$
 \eqalign{
 \bar u^{(1)}_{{\mib k}}& =
\alpha_{\mib k} u^{(1)}_{\mib k}+\beta_{\mib k} u^{*(2)}_{\mib k}\,,
\cr
  \bar u^{(2)}_{{\mib k}}& =
\alpha_{\mib k} u^{(2)}_{\mib k}+\beta_{\mib k} u^{*(1)}_{\mib k}\,,
\cr
}\eqn\bogtwo
$$
and the Euclidean vacuum is given by
$$
|\bar O\rangle={\cal N}\exp\left(\sum_{\mib k}B_{\mib k}
            A^{(1)\dag}_{\mib k}A^{(2)\dag}_{\mib k}\right)
        |O^{(1)}\rangle\otimes|O^{(2)}\rangle.
\eqn\Eucvac
$$

At this point, we note that $\bar u^{(1)}_{\mib k}$
($\bar u^{(2)}_{\mib k}$) is proportional to
$e^{-i\omega_{\mib k}t}$ ($e^{i\omega_{\mib k}t}$)
for the time coordinate $t$ ($=t^{(1)}=t^{(2)}$)
extended over both regions (1) and (2).
This is a result of the fact that, although
regions (1) and (2) are causally disconnected in the Lorentzian
regime, they are analytically connected through the Euclidean regime as
$\tau=it^{(1)}$ and $\tau=\pm\pi+it^{(2)}$.
It is known that the positive (negative) frequency functions
for the Euclidean vacuum are
 characterized by the regularity on the upper-half (lower-half)
complex $t$-plane [\BirDav]. For later convenience, let us
consider the negative frequency functions
(\ie, the analytic continuations at $t=0$ and $t=-i\pi$ through
the lower-half plane).
Then the negative frequency function in the region (1),
$\bar u^{(1)*}_{\mib k}\propto e^{i\omega_{\mib k}t^{(1)}}$,
is analytically continued to the region (2) as
$\bar u^{(1)*}_{\mib k}\propto
 e^{-\omega_{\mib k}\pi+i\omega_{\mib k}t^{(2)}}$.
Similarly, the negative frequency function in the region (2),
$\bar u^{(2)*}_{\mib k}\propto e^{-i\omega_{\mib k}t^{(2)}}$,
is analytically continued to the region (1) as
$\bar u^{(2)*}_{\mib k}\propto
 e^{-\omega_{\mib k}\pi-i\omega_{\mib k}t^{(1)}}$.
Taking the complex conjugates of them, we then find
$$
 \eqalign{
 \bar u^{(1)}_{{\mib k}}& =
\alpha_{\mib k}\left( u^{(1)}_{\mib k}+
 e^{-\omega_{\mib k}\pi}u^{*(2)}_{\mib k}\right),\cr
  \bar u^{(2)}_{{\mib k}}& =
\alpha_{\mib k}\left( u^{(2)}_{\mib k}+
 e^{-\omega_{\mib k}\pi}u^{*(1)}_{\mib k}\right).\cr
}\eqn\anacon
$$
Comparing these with Eq.\bogtwo\ and noting the normalization
condition $|\alpha_{\mib k}|^2-|\beta_{\mib k}|^2=1$, we obtain
$$
 \alpha_{\mib k}={e^{i\theta_{\mib k}}\over
    \sqrt{1-e^{-2\omega_{\mib k}\pi}}}\,,\quad
 \beta_{\mib k}={e^{-\omega_{\mib k}\pi+i\theta_{\mib k}}\over
    \sqrt{1-e^{-2\omega_{\mib k}\pi}}}\,,
\eqn\alpbet
$$
where $\theta_{\mib k}$ is a constant phase.
Also from Eq.\Bkdef, we find
$$
 B_{\mib k}=e^{-\omega_{\mib k}\pi}\,.
\ee
$$

Now it is easy to interpret the above result
to the language of the Schr\"odinger wave functional.
Applying Eq.\Schrvac\ to the present case, we find
$$
     \vert\bar O(t)\rangle_S={\cal N}\exp\left(-{1\over 2\pl}
     \int\int d^3xd^3y~\bar\Omega({\mib x},{\mib y};t)
         \phi({\mib x})\phi({\mib y})\right),
\eqn\Eucschr
$$
where
$$
 \bar\Omega({\mib x},{\mib y};t)
    ={\sqrt{\gamma}\over \alpha}
   \sum_{\mib k}
  \left(\dot{\bar u}^{(1)}_{\mib k}({\mib x},t)
            {\bar u}^{(1)-1}_{\mib k}({\mib y},t)
       +\dot{\bar u}^{(2)}_{\mib k}({\mib x},t)
            {\bar u}^{(2)-1}_{\mib k}({\mib y},t)
     \right).
\ee
$$
However, as we have seen, this state is a zero eigenstate of the
Hamiltonian. Hence the wave functional \Eucschr\ should be
time-independent.
This can be demonstrated as follows.
Operating $e^{-i\hat H_\phi t}$ to Eq.\Eucvac, we obtain
$$
 \vert\bar O(t)\rangle_S=\exp\left(
 \sum_{\mib k} B_{\mib k}
 a^{(1)\dag}_{\mib k}(t^{(1)}) a^{(2)\dag}_{\mib k}(t^{(2)})\right)
 \vert O^{(1)}(t^{(1)})\rangle_S\otimes\vert O^{(2)}(t^{(2)})\rangle_S,
\eqn\bog
$$
where $t^{(1)}=t^{(2)}=t$ and
$a^{(i)\dag}_{\mib k}(t^{(i)})$ are the time-dependent
creation operators as defined in Eq.\Schancr.
 Their coordinate representations are given as
$$
\eqalign{
 a^{(i)}_{\mib k}(t^{(i)}) &
=\int d^3x\left(\pl u^{*(i)}_{\mib k}({\mib x},t^{(i)})
      {\delta\over\delta\phi({\mib x})}
      -i{\sqrt{\gamma}\over\alpha}
        \dot u^{*(i)}_{\mib k}({\mib x},t^{(i)})\phi({\mib x})\right)
\cr
a^{(i)\dag}_{\mib k}(t^{(i)}) &
=\int d^3x\left(-\pl u^{(i)}_{\mib k}({\mib x},t^{(i)})
      {\delta\over\delta\phi({\mib x})}
   +i{\sqrt{\gamma}\over\alpha}
        \dot u^{(i)}_{\mib k}({\mib x},t^{(i)})\phi({\mib x})\right).
\cr}
\ee
$$
Since $\vert O^{(i)}\rangle$ is the true
ground state of the Hamiltonian $\hat H^{(i)}_\phi$,
we have, for $t^{(1)}=t^{(2)}=t$,
$$
 \vert O^{(1)}(t)\rangle_S\otimes \vert O^{(2)}(t)\rangle_S
 = \left(e^{-iE_0 t}\vert O^{(1)}\rangle\right)\otimes
  \left(e^{iE_0 t}\vert O^{(2)}\rangle\right)
= \vert O^{(1)}\rangle\otimes \vert O^{(2)}\rangle.
$$
{}Furthermore, since the Hamiltonian is diagonalized with respect to
the static vacuum mode functions, we have
$$
 a^{(1)\dag}_{\mib k}(t) a^{(2)\dag}_{\mib k}(t)
=\left(e^{-i\omega_{\mib k}t}A^{(1)\dag}_{\mib k}\right)
 \left(e^{i\omega_{\mib k}t}A^{(2)\dag}_{\mib k}\right)
=A^{(1)\dag}_{\mib k}A^{(2)\dag}_{\mib k}\,.
\ee
$$
Therefore we find
$$
 \vert\bar O(t)\rangle_S=\exp\left(
 \sum_{\mib k} B_{\mib k} A^{(1)\dag}_{\mib k}
               A^{(2)\dag}_{\mib k}\right)
 \vert O^{(1)}\rangle\otimes\vert O^{(2)}\rangle.
\eqn\superposition
$$
This form is explicitly time-independent.

Now, as we have obtained the asymptotic behavior of
the wave functional in the false vacuum,
we go back to the problem of constructing
the tunneling wave functional.
As clear from Eq.\superposition,
the Euclidean vacuum can be considered as a
superposition of many particle states of the following form,
$$
 (A^{(1)\dag}_{{\mib k}_1})^{n_1}
 \cdots(A^{(1)\dag}_{{\mib k}_m})^{n_m}
  \vert O^{(1)}\rangle\otimes
 (A^{(2)\dag}_{{\mib k}_1})^{n_1}
 \cdots(A^{(2)\dag}_{{\mib k}_m})^{n_m}
  \vert O^{(2)}\rangle.
\eqn\asym
$$
As there is no causal connection between the regions (1) and (2),
the wave functional can be constructed independently in each
region. First consider the region (1).
Applying the results obtained in the previous section,
the wave functional for the region (1) which has the asymptotic form
in the false vacuum as
$$
 (A^{(1)\dag}_{{\mib k}_1})^{n_1}
\cdots(A^{(1)\dag}_{{\mib k}_m})^{n_m}
  \vert O^{(1)}\rangle
\ee
$$
is given by
$$
 (A^{(1)\dag}_{{\mib k}_1}(t^{(1)}))^{n_1}
\cdots(A^{(1)\dag}_{{\mib k}_m}(t^{(1)}))^{n_m}
  \vert O^{(1)}(t^{(1)})\rangle
\eqn\asym
$$
where $|O^{(1)}(t^{(1)})\rangle$ is the quasi-ground-state
wave functional with the parameter time $t^{(1)}$
along the path shown in Fig.~2.
The operators $A^{(1)}_{\mib k}(t^{(1)})$ and
 $A^{(1)\dag}_{\mib k}(t^{(1)})$ are
given by
$$
\eqalign{
  A^{(1)}_{{\mib k}}(t^{(1)})
 & = e^{-i\omega_{\mib k}t^{(1)}}a^{(1)}_{\mib k}(t^{(1)})
\cr
 & =e^{-i\omega_{\mib k}t^{(1)}}
  \int d^3x  \left(\pl w^{(1)}_{\mib k}({\mib x},t^{(1)})
     {\delta\over\delta\phi({\mib x})}
-i{\sqrt{\gamma}\over\alpha}
    \dot w^{(1)}_{\mib k}({\mib x},t^{(1)})\phi({\mib x})\right),
\cr
 A^{(1)\dag}_{\mib k}(t^{(1)})
& = e^{i\omega_{\mib k}t^{(1)}}a^{(1)\dag}_{\mib k}(t^{(1)})
\cr
 &= e^{i\omega_{\mib k}t^{(1)}}
\int d^3x \left(-\pl v^{(1)}_{\mib k}({\mib x},t^{(1)})
 {\delta\over\delta\phi({\mib x})}
 +i{\sqrt{\gamma}\over\alpha}
 \dot v^{(1)}_{\mib k}({\mib x},t^{(1)})\phi({\mib x})\right),
\cr
}\ee
$$
where $v^{(1)}_{\mib k}$ and $w^{(1)}_{\mib k}$
satisfy the field equation,
$$
 \left[g^{\mu\nu}\nabla_{\mu}\nabla_{\nu}
 -\left(m^2(\sigma)+\xi R\right)\right] z^{(1)}_{\mib k}=0\quad
(z=v, w),
\eqn\eumode
$$
along the trajectory on the complex $t^{(1)}$-plane
shown in Fig.~2 with the asymptotic
initial condition at $t^{(1)}<0$,
$$
\eqalign{
 v^{(1)}_{\mib k}({\mib x},t^{(1)})&
  =u^{(1)}_{\mib k}({\mib x},t^{(1)}),
\cr
 w^{(1)}_{\mib k}({\mib x},t^{(1)})&
  =u^{(1)*}_{\mib k}({\mib x},t^{(1)}).\cr
}\ee
$$
As for the wave functional for the region (2),
since the tunneling degree of freedom
keeps staying at the false vacuum origin, it is $t^{(2)}$-independent.
However, we may express the wave functional in the same manner
as that for region (1) by solving the mode functions
$v^{(2)}_{\mib k}$ and $w^{(2)}_{\mib k}$ along
the contour of Fig.~2 on the complex $t^{(2)}$-plane, but
with the false vacuum configuration throughout the contour.

Summing up all the terms again, we obtain the wave functional
which describes tunneling from the Euclidean vacuum,
$$
 \Phi\left[\phi(\,\cdot\,);t\right]=
 {\cal N} \exp\left(
 \sum_{\mib k} B_{\mib k} A^{(1)\dag}_{\mib k}(t)
                 A^{(2)\dag}_{\mib k}\right)
 \vert O^{(1)}(t)\rangle\otimes\vert O^{(2)}\rangle.
\eqn\bbb
$$

Now, as we have discussed when deriving the Bogoliubov
coefficients \alpbet, when we consider the negative frequency functions,
the regions (1) and (2) are analytically
connected through the lower-half complex $t$-plane
while the tunneling field is at the false vacuum origin.
Hence we expect the tunneling wave functional to be obtained
by finding the mode functions $\bar w^{(i)}_{\mib k}$ for
the tunneling background which correspond to the Euclidean vacuum
negative frequency functions $\bar u^{(i)*}_{\mib k}$.
Namely, first we set the boundary condition,
$$
\bar w^{(i)}_{\mib k}({\mib x},t)=\bar u^{(i)*}_{\mib k}({\mib x},t)\,,
\ee
$$
at $t<0$. Then they solve the field equation \eumode\
in the Lorentzian time to $t=t^{(1)}=t^{(2)}=0$, and further in the
Euclidean time with $\tau=it^{(1)}$ and $\tau=-\pi+it^{(2)}$
 beyond $\tau=-\pi$ to $\tau=-\infty$.
Note that from Eq.\anacon, the solutions are automatically
consistent with the analytic continuation at both $\tau=0$ and $-\pi$.
Then solving back to $\tau=0$ through
the non-trivial $O(4)$ bubble background only in the interval
$-\pi/2<\tau\leq0$, and analytically continue to the Lorentzian
time with $t=t^{(2)}=-i(\tau+\pi)$ at $\tau=-\pi$ to
the de Sitter background
and $t=t^{(1)}=-i\tau$ at $\tau=0$ to the $O(3,1)$ bubble background.

Then, with thus obtained mode functions $\bar w^{(i)}_{\mib k}$,
we can in fact show that the tunneling wave functional satisfies
$$
 \bar a^{(i)}_{{\mib k}}(t) \Phi\left[\phi(\,\cdot\,);t\right]=0,
\ee
$$
where
$$
 \bar a^{(i)}_{{\mib k}}(t)
  =\int d^3x
  \left(\pl\bar w^{(i)}_{\mib k}({\mib x},t)
        {\delta\over\delta\phi({\mib x})}
 -i{\sqrt{\gamma}\over\alpha}
 \dot{\bar w}{}^{(i)}_{\mib k}({\mib x},t)\phi({\mib x})\right).
\ee
$$
After all, as clear from the above procedure, we do not have to know
$\bar w^{(i)}_{\mib k}$ at $\tau\leq-\pi$ but only those
in the interval $-\pi\leq\tau\leq0$. Furthermore
 $\bar w^{(i)}_{\mib k}$ coincide with $\bar u^{(i)}_{\mib k}$
in the region $\tau\leq-\pi/2$.
Once we become aware of these facts, we need not stick to the
construction of $\bar w^{(i)}_{\mib k}$ themselves, nor to
the coordinates of the metric \static.
Instead, any complete set of
mode functions which are related to $\bar w^{(i)}_{\mib k}$ by
a unitary transformation is relevant and any convenient
coordinate system can be chosen to solve for them. For this
 reason, we may drop the superscript $(i)$ for the
mode functions and denote them simply by $\bar w_{\mib k}$.
Thus the procedure to construct the tunneling wave functional
in the case with gravity
turns out to be very similar to the case without gravity.
In particular, the resulting quantum state after tunneling will
be related to the true vacuum state by a Bogoliubov transformation
as discussed in Papers I and II.

\section{Conformal scalar model}

Here, as a simple application of our formalism, we
consider a conformally coupled scalar field $\phi$
(\ie, $\xi=1/6$ in Eq.\hHel)
which is massless except on the bubble wall.
In Paper II, we have investigated a similar model in the absence
of the background curvature. We show below that this conformal
scalar model gives the quantum state which is conformally equivalent
 to the one without gravity in Paper II.

A convenient choice of the coordinates for the present case
is obtained by the coordinate transformation of
 the metric \metric\ with $n(\eta)=a(\eta)$ as
$$
 \eqalign{
 T_E & =-e^\eta\cos r,\cr
 \rho & =e^\eta\sin r,\cr
}\ee
$$
or equivalently that of the metric \static\ as
$$
\eqalign{
T_E=&{\sqrt{1-R^2}\sin\tau(1+\sqrt{1-R^2}\cos\tau)\over
         \sin^2\tau+R^2\cos^2\tau}\,,
\cr
\rho&={R(1+\sqrt{1-R^2}\cos\tau)\over
         \sin^2\tau+R^2\cos^2\tau}\,.
\cr
}\eqn\terho
$$
Then the metric becomes
$$
 ds_E^2=\Omega_E^2(\xi_E)
    \left(dT_E^2+d\rho^2+\rho^2 d\Omega^2_{(2)}\right),
\eqn\Emetric
$$
where
$$
\Omega_E^2={a^2(\ln\xi_E)\over \xi_E^2}\,;\quad
\xi_E=\sqrt{T_E^2+\rho^2}\,.
\ee
$$
As clear from Eq.\terho, the analytic continuation to the
Lorentzian metric by $\tau=it$ ($t=t^{(1)}$) and
 $\tau=-\pi+it$ ($t=t^{(2)}$) corresponds to that
at $T_E=0$ by $T_E=iT$.
Hence the Lorentzian version of the metric \Emetric\ is
$$
 ds^2=\Omega^2(\xi)
    \left(-dT^2+d\rho^2+\rho^2 d\Omega^2_{(2)}\right),
\eqn\Lmetric
$$
where
$$
\Omega^2={a^2(\ln\xi)\over \xi^2}\,;\quad
\xi=\sqrt{-T^2+\rho^2}\,.
\eqn\confac
$$
At the false vacuum origin, Eq.\Lmetric\ reduces to the
de Sitter metric:
$$
\eqalign{
 ds^2_{deS}&=\Omega_{deS}^2(\xi)
    \left(-dT^2+d\rho^2+\rho^2 d\Omega^2_{(2)}\right);
\cr
&\Omega_{deS}={2\over H_F(1-T^2+\rho^2)}\,.
\cr}
\eqn\OmedeS
$$

As we have seen in the previous subsection,
the procedure to obtain the tunneling wave functional
is to solve for the mode functions $\bar w_{\mib k}$ in the interval
$-\pi\leq\tau\leq0$ of the Euclidean tunneling background
and analytically continue them to the Lorentzian background
at $\tau=-\pi$ and $\tau=0$ with the condition that
$\bar w_{\mib k}$ are unitarily equivalent to the analytic continuation
of the Euclidean vacuum negative frequency functions
$\bar u^*_{\mib k}$ in the region $\tau\leq-\pi/2$.
Hence we may regard Eq.\Lmetric\ to represent
the background metric both in the Euclidean and Lorentzian regimes
by allowing $T$ to take complex values.
We show how the time coordinate $T$ spans
the spacetime in Fig.~6.
The advantage of using the coordinates $(T,\rho)$ is that
the correspondence to the flat spacetime case becomes transparent.
The metric \Lmetric\ is conformally equivalent to the flat spacetime
metric: $g_{\mu\nu}=\Omega^2\eta_{\mu\nu}$. Further,
the conformal factor $\Omega$ is a function of only $\xi$, \ie,
it is $O(3,1)$- (or $O(4)$-) invariant.
In particular, $\Omega$ may be regarded as a function of
the tunneling field $\sigma$.

Because of the conformal coupling of the scalar field,
the field equation \eumode\ for the mode functions
 $\bar w_{\mib k}$ can be conformally transformed
to that on the flat spacetime:
$$
 \left[\eta^{\mu\nu}\nabla_{\mu}\nabla_{\nu}
 -\left(m^2(\sigma)\Omega^2(\sigma)\right)\right]
 \bar w_{f\mib k}=0,
\eqn\flatmode
$$
where we have regarded $\Omega$ as a function of $\sigma$ and
$$
 \bar w_{f\mib k}=\Omega\bar w_{\mib k}\,.
\eqn\wfdef
$$
Thus we can construct the mode functions which satisfy Eq.\eumode\
 by solving Eq.\flatmode\ in the flat spacetime.

The remaining task is to impose the correct boundary condition
on $\bar w_{f\mib k}$. For this purpose,
let us consider the case of pure de Sitter background.
If the conformal vacuum defined by the positive frequency
functions $u_{f\mib k}$ ($\propto e^{-ikT}$)
on the flat space agrees with the Euclidean vacuum,
the required initial condition for $\bar w_{f\mib k}$
is trivial; $\bar w_{f\mib k}=u^*_{f\mib k}$.
To see this is indeed the case, we examine the symmetric two-point
function in the conformal vacuum.
It is well-known that the positive frequency function $\bar u_{\mib k}$
for this conformal vacuum is given by
 $\bar u_{\mib k}=\Omega^{-1}\bar u_{f\mib k}$
and as a result the two-point function $G^{(1)}$ is given by
$$
 G^{(1)}(x,x') =\Omega_{deS}^{-1}(x^2)D^{(1)}(x,x')
        \Omega_{deS}^{-1}({x'}^2),
\eqn\confG
$$
where $\Omega_{deS}$ is given by Eq.\OmedeS,
$x^2=-(x^0)^2+{\mib x}^2=-T^2+\rho^2$ and
$$
 D^{(1)}(x,x'):={1\over 2\pi^2}{1\over (x-x')^2},
\ee
$$
is the symmetric two-point function in the Minkowski vacuum.

If we embed the de Sitter space in
the five dimensional Minkowski space,
$$
 ds^2=-(d\tilde x^0)^2+(d\tilde x^1)^2+(d\tilde x^2)^2
+(d\tilde x^3)^2+(d\tilde x^4)^2,
\ee
$$
we find the coordinates $\tilde x^a$ $(a=0,1,2,3,4)$
on the de Sitter space are expressed in terms of
 $x^\mu=(T,{\mib x})$ as
$$
 (\tilde x^\mu,\tilde x^4)=
 \left({2x^\mu\over H_F(1+x^2)}\,,{1-x^2\over H_F(1+x^2)}\right),
\ee
$$
It is then easy to show that
$$
 G^{(1)}(x,x')={1\over 2\pi^2}{1\over (\tilde x-\tilde x')^2}.
\ee
$$
This shows $G^{(1)}$ is de Sitter invariant.
Hence, together with the fact that
its short distance behavior is the same as that
in the Minkowski vacuum, the present conformal vacuum is
found to be the Euclidean vacuum.
The relevant mode functions $\bar w_{\mib k}$ for the
tunneling wave functional are then given by solving
Eq.\flatmode\ along the contour shown in Fig.~2 on the complex
$T$-plane and multiplying the result by the inverse of the
conformal factor; $\bar w_{\mib k}=\Omega^{-1}\bar w_{f\mib k}$.

To summarize, in the present case of a
conformally coupled scalar field, the effects of
gravity to the quantum state after tunneling
is to solve the mode functions for the flat background
with the mass term $m^2$ replaced by $m^2\Omega^2$
and multiplying the resultant
mode functions by $\Omega^{-1}$, with $\Omega$ given in Eq.\confac.
The two-point functions for the quantum state after tunneling
are also given by the conformal transformation of those
obtained for the flat background $D(x,x')$ as
$$
 G(x,x') =\Omega^{-1}(x^2)D(x,x')\Omega^{-1}({x'}^2).
\eqn\confGG
$$

However, the evaluation of the energy momentum tensor
seems to require some care. It is known that the regularized
vacuum expectation value of the energy momentum tensor,
$\VEV{T^{\mu\nu}}$, for a conformally
coupled field on a conformally flat spacetime consists
of term arising from a trivial conformal transformation
of $\VEV{T_f^{\mu\nu}}_{reg}$ in flat space
and the terms representing conformal anomalies [\Anomaly]:
$$
\VEV{T_{\mu\nu}}_{reg}=\Omega^{-2}\VEV{T^f_{\mu\nu}}
   +\hbox{\rm conformal anomalies}.
\eqn\Tmunu
$$
It is also known that if there arises no further
divergence apart from the common ones for any spacetime,
Eq.\Tmunu\ continues to hold for arbitrary state.
However, in the present case, as discussed in Paper II,
we encounter a new type of divergences which may be partly
due to the delta-function nature of the mass term in our model
and also possibly due to the breakdown of the WKB expansion.
Unfortunately at the moment, we are unable to clarify if
these new divergences would give rise to terms which are
conformally non-trivial.
If not, since the finite terms
of $\VEV{T^f_{\mu\nu}}$ we have found in flat space in Paper II will
be absent in the case of conformal coupling, except for the term
which diverges on the light cone,
we would find only the conformal anomaly terms in the present model,
provided we impose the regularity on the light cone. Furthermore,
if the true vacuum has no vacuum energy, it reduces
to the flat space and we would find no finite term at all,
which sounds rather paradoxical.
In Paper II, we have argued that the regularity on the light cone
is necessary to keep the validity of the WKB expansion.
Hence if we allow the presence of the term which
diverges on the light cone, it will be necessary to seriously
consider the possible breakdown of the WKB expansion.
The resolution of this issue is left for future study.

\chapter{Conclusions}

We have considered an extension of our previous analysis in Paper II
of the quantum state after $O(4)$-symmetric bubble nucleation in flat
space to the case with the gravitational effect.
In order to do so, we have first extended the formalism developed in
Paper I to the case of multi-dimensional tunneling from an
excited state at the false vacuum origin.
Then using the result of extension, we have developed a method
to obtain the tunneling wave functional from the false vacuum
to the true vacuum through a non-trivial geometry of the
background spacetime described by the $O(4)$-symmetric bubble
with gravity. Provided that the $O(4)$ bubble is described by
the thin-wall approximation, we have found the procedure to
 construct the tunneling wave functional can be formulated
in a quite similar manner as
in the case of flat spacetime background.

As an explicit demonstration of our formalism, we have
considered a simple conformal scalar model which is massless
except on the bubble wall to represent the fluctuations around
the $O(4)$ bubble. We have then found the
the resulting quantum state is conformally equivalent to
that in the absence of gravity, \ie, it is described by
a Bogoliubov transformation of the true vacuum state
(a squeezed state).
However, we have argued that the evaluation of the regularized
expectation value of the energy momentum tensor for this quantum
state may be highly non-trivial, apart from the conventional conformal
anomalies.
We have also pointed out the paradoxical situation that
 the regularized energy momentum tensor might vanish
 due to the conformal coupling nature of our model
if the true vacuum has no vacuum energy density.

At the moment, we are unfortunately unable to judge whether these
issues are particularities of our over-simplified model
or intrinsic difficulties associated with field theoretical
tunneling phenomena. Further research on the present subject
is apparently required.

\ack{
This work was supported by Monbusho Grant-in-Aid for
 Scientific Research
Nos. 2010 and 05640342, and the Sumitomo Foundation.}

\par \penalty-400 \vskip\chapterskip
   \spacecheck\referenceminspace \immediate\closeout\referencewrite
   \referenceopenfalse
   \line{\fourteenrm\hfil REFERENCES\hfil}\vskip\headskip
   \input reference.aux
   
\vskip2cm
\centerline{\seventeenrm Figure Captions}

\item{\bf Fig.~1:} The potential form of the tunneling field, where
$\sigma_F$ and $\sigma_T$ represent the values of
$\sigma$ in the false and true vacua, respectively.

\item{\bf Fig.~2:} A path on the complex plane of time,
which represents a tunneling process.
The segments A, B and C correspond to the motion staying in the
false vacuum, an instanton and the motion after nucleation,
respectively.

\item{\bf Fig.~3:} A schematic picture of the
Coleman DeLuccia instanton solution in the thin wall limit.
Two dimensions are suppressed

\item{\bf Fig.~4:} Foliation of the Coleman DeLuccia
instanton solution by $\tau=$ const. hypersurfaces.

\item{\bf Fig.~5:} The analytic continuation
to the Lorentzian region of the Coleman DeLuccia instanton solution.
The lower and upper halves are Euclidean and Lorentzian regions,
respectively.

\item{\bf Fig.~6:} The same as Fig.~5, but with foliation by
 the $T=$ constant hypersurfaces.

\end